# Substitutional VS$_n$ Nanodispersed in MoS$_2$ Film for Pt-scalable Catalyst

*Frederick Osei-Tutu Agyapong-Fordjour, Seok Joon Yun, Hyung-Jin Kim, Wooseon Choi, Soo Ho Choi, Laud Anim Adofo, Stephen Boandoh, Yong In Kim, Soo Min Kim, Young-Min Kim, Young Hee Lee\*, Young-Kyu Han\*, and Ki Kang Kim\**

F. O. T. Agyapong-Fordjour, W. Choi, L. A. Adofo, Y. I. Kim, Prof. Y.-M Kim, Prof. Y. H. Lee, Prof. K. K. Kim
Department of Energy Science, Sungkyunkwan University, Suwon 16419, Republic of Korea.

Dr. S. J. Yun, Dr. S. H. Choi, Dr. S. Boandoh, Prof. Y.-M Kim, Prof. Y. H. Lee, Prof. K. K. Kim
Center for Integrated Nanostructure Physics (CINAP), Institute for Basic Science (IBS), Sungkyunkwan University, Suwon 16419, Republic of Korea.

F. O. T. Agyapong-Fordjour, Dr. H. J. Kim, Prof. Y. K. Han
Department of Energy and Materials Engineering, Dongguk University, Seoul 04620, Republic of Korea.

Prof. S. M. Kim
Department of Chemistry, Sookmyung Women's University, Seoul 140742, Republic of Korea.

E-mail: leeyoung@skku.edu, ykenergy@dongguk.edu and kikangkim@skku.edu

Among transition metal dichalcogenides (TMdCs) as alternatives for Pt-based catalysts, metallic-TMdCs catalysts have highly reactive basal-plane but are unstable. Meanwhile, chemically stable semiconducting-TMdCs show limiting catalytic activity due to their inactive basal-plane. Here, we propose metallic vanadium sulfide (VS$_n$) nanodispersed in a semiconducting MoS$_2$ film (V-MoS$_2$) as an efficient catalyst. During synthesis, vanadium atoms are substituted into hexagonal monolayer MoS$_2$ to form randomly distributed VS$_n$ units. The V-MoS$_2$ film on a Cu electrode exhibits Pt-scalable catalytic performance; current density of 1000 mA cm$^{-2}$ at 0.6 V, overpotential of -0.06 V at a current density of 10 mA cm$^{-2}$ and exchange current density of 0.65 mA cm$^{-2}$ at 0 V with excellent cycle stability for hydrogen-evolution-reaction (HER). The high intrinsic HER performance of V-MoS$_2$ is explained by the efficient electron transfer from the Cu electrode to chalcogen vacancies near vanadium sites with optimal Gibbs free energy (-0.02 eV). This study adds insight into ways to engineer TMdCs at the atomic-level to boost intrinsic catalytic activity for hydrogen evolution.

One renewable energy resource is water electrolysis for hydrogen production. Noble metals such as Pt and Pd as catalysts for water electrolysis are inevitably utilized for efficient hydrogen evolution,[1-3] which have been limited for industry application owing to scarcity and high cost. Therefore, it is highly desired to develop efficient noble-metal free catalyst. Among a variety of noble-metal free catalysts, TMdC-layered material has been proposed. Metallic TMdCs such as VS$_2$,[4] VSe$_2$[5] and NbS$_2$[6] demonstrate great potential for HER performance owing to their active basal plane and high electrical conductivity but are often not stable in air,[7-9] and most importantly in HER environment,[5,10] which is a primary concern for industrial targets. Meanwhile, semiconducting TMdCs such as MoS$_2$ and WS$_2$ in monolayer form[11-13] are stable in air but are inactive in the basal plane and have poor electrical conductivity. Additionally, monolayer TMdCs allow for short electron injection paths from the electrode to active sites to promote efficient HER performance, unlike the bulk materials, which offers limited HER kinetics because of charge lagging to reach active surface sites.[14,15]

Several approaches with semiconducting TMdCs have been investigated to resolve the basal-plane inactivity including chalcogen vacancies, phase boundaries, and heterostructures.[16-18] Although the hydrogen adsorption Gibbs free energy ($\Delta G_{H^*}$) approaches a nearly zero in chalcogen vacancies and heterostructures, electrical conductivity remains poor, leaving sluggish HER kinetics.[14,15,18] Heterophase boundaries are highly reactive in terms of HER but the necessary phase-boundary line density is rarely encountered.[17] Structural transformation from the semiconducting 2H-phase to metallic 1T phase renders these materials reactive with respect to HER performance but unfavorable owing to the thermodynamically unstable 1T phase.[19] Recently, the substitution of metal and chalcogen atoms in bulk s-TMdCs prepared by hydrothermal method improved the HER performance, but the complicated structures make it difficult to understand the underlying HER mechanism.[20-22] Our research target is to take advantage of the highly conductive basal plane of monolayer TMdC film in large-area for efficient HER performance, yet to obtain a material with high stability. Here we construct a one-step chemical vapor deposition (CVD) process to directly synthesize metallic vanadium sulfide (VS$_n$) units nanodispersed in semiconducting monolayer MoS$_2$ film (V-MoS$_2$) and further demonstrate superb HER performance *via* Tafel slope, overpotential, exchange current density, turnover frequency, cycle test, Gibbs free energy that are comparable to those of Pt catalyst.

**Figure 1**a depicts the ball-and-stick model for randomly dispersed $VS_n$ unit in the active basal plane of a hexagonal $MoS_2$ monolayer. Most importantly, the chalcogen vacancy is created near the vanadium site, which is defined as a $VS_n$ unit. This unit plays an important role as active sites in the basal plane to promote both efficient charge transfer and hydrogen adsorption. The nanodispersed $VS_n$ units in monolayer $MoS_2$ film was synthesized by one-step CVD process (see Experimental Section and Figure S1, Supporting Information). Atomic structures and the homogeneity of V distribution in V-$MoS_2$ at 9.3 at% V concentration are provided in an annular dark field scanning transmission electron microscopy (ADF-STEM) image (Figure 1b, Figures S2 and S3, Supporting Information). The d-spacing between S-S in the $VS_n$ unit, which is confirmed by observing the $(10\bar{1}0)$ plane from the electron diffraction pattern shown in the inset, is 0.27 nm, similar to that of pristine 2H-$MoS_2$ (0.27 nm).[23] The real and simulated ADF-STEM images (the square region in Figure 1b) show the atomic configuration of the $VS_n$ unit region with an S vacancy in V-$MoS_2$ (Figure 1c). Furthermore, the intensity profile clearly distinguishes among Mo, S, and V atoms and the S vacancy next to the V atom (V-$vac_s$) (Figure 1d and inset: top view of the atomic configuration). These results indicate that the V atoms are well substituted into Mo sites in the 2H-$MoS_2$ lattice with negligible strains. The formation of $VS_n$ units was readily confirmed by newly emerged peak in Raman spectra and reduced photoluminescence intensity due to the enhanced metallic character (Figures S1, Supporting Information). The presence of V atoms in the as-grown V-$MoS_2$ film was additionally confirmed by energy dispersive X-ray spectroscopy and X-ray photoelectron spectroscopy (Figures S4-S6, Supporting Information).

We carefully analyzed the atomic configurations of V at the Mo site ($V_{Mo}$), V-$vac_s$, Mo at the Mo site ($Mo_{Mo}$), the S vacancy next to the Mo atom (Mo-$vac_s$), and the S dimer (2S) (Figure 1e and Figure S7, Supporting Information). Interestingly, the amount of V-$vac_s$ is proportional to the $V_{Mo}$ concentration, the molar ratio of V to the Mo precursor (Figure 1f and Figures S8-S12, Supporting Information for more details), with a similar trend to that of the longitudinal acoustic mode phonon in Raman spectra originating from impurity levels[24] (Figure S13, Supporting Information). The coverage of monolayer region reaches up to 95 % (5 % multilayer region) in $V_{(9.3\%)}$-$MoS_2$ (Figure S14 and S15, Supporting Information). Meanwhile, the amount of Mo-$vac_s$ in V-$MoS_2$ is stochastically populated up to 2.1%, regardless of $V_{Mo}$ amount (Figure 1f). The higher V-$vac_s$ amount of ~22.0% per V atom than Mo-$Vac_s$ amount of 1.2% per Mo atom is congruent with our theoretical calculations, where the formation energy of V-vac$_s$ is 0.22 eV more stable than that of Mo-vac$_s$ (Table S1, Supporting Information). The ample V$_{Mo}$ and V-vac$_s$ sites are the key ingredient for efficient electron transfer and hydrogen adsorption, which will be discussed later.

**Figure 2**a shows the polarization curves in the linear sweep voltammetry plot with varying V concentration and substrate (see Experimental Section for a detailed description of the measurements). Ni and Cu electrodes were chosen for their earth abundance and relatively high electrical conductivities in comparison with conventional graphite (Gr). The polarization curve of V-MoS$_2$ with the Gr substrate shifts to the Pt curve with increasing V concentration, surpassing that of pristine MoS$_2$. The overpotential with the Ni substrate is further reduced relative to that with the Gr substrate. The Cu substrate with 9.3 at% V reveals the lowest overpotential, closely approximating that of Pt. The Tafel slope in the presence of V content exceeds that of pristine MoS$_2$, decreasing to 54.2 mV dec$^{-1}$ in V$_{(9.3\%)}$-MoS$_2$ with the Gr substrate (Figure 2b). This value is reduced to 46 mV dec$^{-1}$ with the Ni substrate and further to 40 mV dec$^{-1}$ with the Cu substrate. It is remarkable to see that with the Cu substrate, V-MoS$_2$ show a similar overpotential in low current density region to that of Pt.

The overpotential for 10 mA cm$^{-2}$ ($\eta_{10}$) and exchange current density ($j_0$) for assessing catalytic efficiency are extracted from linear sweep voltammetry plot and Tafel equation.[25] The overpotential gradually decreases to -120 mV in V$_{(9.3\%)}$-MoS$_2$ with the Gr substrate, whereas $j_0$ slowly increases to 0.283 mA cm$^{-2}$ (Figure 2c). Both $\eta_{10}$ and $j_0$ are further improved by using a different substrate. In particular, V$_{(9.3\%)}$-MoS$_2$ with the a Cu substrate demonstrates the lowest $\eta_{10}$ of -0.06 V and highest $j_0$ of 0.65 mA cm$^{-2}$ corresponding to hydrogen turnover-frequency (TOF) of 0.3 s$^{-1}$ (Figure 2d), comparable to those of Pt[26] ($\eta_{10}$ of -0.02 V, $j_0$ of 0.8 mA cm$^{-2}$ and TOF of 0.7 s$^{-1}$). Furthermore, the current density reaches to 1000 mA cm$^{-2}$ at 0.6 V, which is viable in industrial requirements (Figure S16, Supporting Information). The outstanding onset potential, $\eta_{10}$, Tafel slope, and $j_0$ of V$_{(9.3\%)}$-MoS$_2$ are further compared with those of other 2D materials (Figure S17, Supporting Information).

To evaluate the stability in HER environment, we perform chronoamperometry measurements and an accelerated degradation test. The V$_{(9.3\%)}$-MoS$_2$ sample clearly demonstrates no significant drop in the current density at different current densities of 1, 10, and 50 mA cm$^{-2}$ for 24 hours (Figure 2e). Additionally, the polarization curves are fully superimposed after 5000 cycles, regardless of the substrate used (Figure 2f). Also, the chemical shifts, the change of morphology, and physical structure before and after cycling were negligible (XPS, SEM, and Raman analysis in Figure S18, Supporting Information). This superb stability is crucial to meet industrial target and is well contrasted with that of pure metallic $VS_2$.[4]

To investigate the origin of the charge transfer kinetics and catalytically active sites in V-$MoS_2$, we measured the charge-transfer resistance ($R_{ct}$) and double-layer capacitance ($C_{dl}$) by using impedance spectroscopy and cyclic voltammetry, respectively (see Experimental Section and Figures S19-S21, Supporting Information). The charge transfer resistance rapidly drops at the minute concentration of 0.8 at% V from that of pristine $MoS_2$ and gradually reduces to saturate at a concentration of 9.3 at% V (Figure 2g). Such a drastic reduction in $R_{ct}$ is ascribed to the degenerate metallic $VS_2$ of the high concentration of 9.3 at% V in the semiconducting $MoS_2$ lattice.[27] The $R_{ct}$ for $V_{(9.3\%)}$-$MoS_2$/Cu is reduced to 0.8 Ω, which is the lowest $R_{ct}$ value ever recorded for a 2D monolayer TMdC electrocatalyst to date, and even lower than that of Pt (2.2 Ω) (Figure S19, Supporting Information). The electrochemically active surface area (EASA) extracted from the $C_{dl}$[28] is nearly negligible for pristine $MoS_2$ compared to that of Gr (Figure S22 and Table S2, Supporting Information.). This value is gradually elevated with the V concentration, reaching to 28.2 mF cm$^{-2}$ for $V_{(9.3\%)}$-$MoS_2$, twice as high as that of pristine $MoS_2$ (13.2 mF cm$^{-2}$). The EASA with the Cu substrate in $V_{(9.3\%)}$-$MoS_2$ is further improved from the Gr substrate, ensuring the importance of substrate for industrial applications. Furthermore, EASA normalized polarization curves demonstrate outstanding HER performance of V-$MoS_2$/Cu as compared to other materials (Figure S23 and Table S3, Supporting Information).

Since the catalytically active sites are directly related to EASA, we elucidate the underlying mechanism on active sites in V-$MoS_2$ by performing density functional theory (DFT) calculations. Here, we focus on two main aspects: the Gibbs free energy ($\Delta G_{H^*}$) with substrates and density of states (DOS) near the Fermi level associated with active sites. Typical chalcogen vacancies on pure $MoS_2$ and V-$MoS_2$ and various substrates including Gr (0002), Cu (111), and Ni (111) are schematically drawn in **Figure 3**a. The Gibbs free energy at chalcogen vacancy next to Mo sites (Mo-$vac_s$) shows relatively close to the thermoneutral point, although its dependence on the substrate varies slightly (Figure 3b). It is remarkable to reveal the best Gibbs free energy at chalcogen vacancy next to V site (V-$vac_s$) on the Cu substrate is -0.02 eV, nearly ideal 0 eV, while the other substrates (Gr and Ni) are far deviated from the thermoneutral point. The volcano plot is summarized with various materials in the literature (Figure 3c, Figures S24 and S25, Supporting Information). While the Gibbs free energy of Mo-$vac_s$ is similar to that of V-vac$_s$ with Cu substrate, the exchange current density of V-vac$_s$ is superior to that of Mo-vac$_s$ (Table S4, Supporting Information). A summary of catalytic parameters of V-MoS$_2$/Cu in comparison with those of Pt and other TMdCs catalysts are provided in **Table 1**.

The charge-transfer to the active site is directly related to the DOS near the Fermi level.[6,35] We consider the S vacancies with pristine MoS$_2$, V-MoS$_2$ with two V concentrations, and VS$_2$ on Cu substrate (Figure 3d) and calculate projected DOS (PDOS) of individual atoms in (5x5) unit cell (see Experimental Section). The PDOS of MoS$_2$ with a single S vacancy (Figure 3e) is developed near the Fermi level (E = E$_F$) and is further developed with additional V site in V$_{(4\%)}$-MoS$_2$/Cu. The bandgap is still preserved at this V concentration. In addition to enhanced DOS at V$_{(16\%)}$-MoS$_2$/Cu, the bands are highly degenerate. The bandgap is closed in metallic VS$_2$/Cu and therefore the stability of material is no longer guaranteed. The integrated DOS near the Fermi level are elevated with increasing V concentrations and promotes the electron injection to active sites. The higher V concentration, the better the exchange current density but limited by material stability like pure VS$_2$ (Figure 3f).

The nanodispersed VS$_n$ in the semiconducting MoS$_2$ lattice in our study is certainly advantageous in several respects (Figure 3g). The content of chalcogen vacancies in V-MoS$_2$ is ~1.0 x 10$^{14}$ cm$^{-2}$ at 9.3 at% V concentration, which is about twice the amount of S vacancies in pristine MoS$_2$.[35] Consequently, the exchange current density meets the industrial target. An additional advantage is that the substrate is highly susceptible to proximate monolayer V-MoS$_2$, which can be used to engineer Gibbs free energy and electron injection. In particular, the Cu substrate is not only useful to improve the exchange current density but also to tune the Gibbs free energy to the ideal 0 eV, facilitating efficient electron transfer from the Cu substrate to active sites. Another engineering point is preserving the material stability. In our case, the unstable VS$_2$ metal is stabilized by introducing nanodispersed VS$_n$ in the semiconducting MoS$_2$ lattice. At this stage, raising the V concentration beyond 10% V is limited from a synthesis point of view due to incomplete formation of fully covered V-MoS$_2$ film (Figure S26, Supporting Information). This is an open question to increase further the V- concentration in order to improve catalyst efficiency, while ensuring stability. The exchange current density can be further improved by introducing 3D scaffolds such as wrinkles or a porous network. Our strategy provides insight into ways to engineer a single-atom catalyst at the atomic level with 2D materials and furthermore facilitates the design of target-specific characteristics for application to a variety of electrocatalysts, photocatalysts, and electronic devices.

**Experimental Section**

*Growth of MoS$_2$ and V-MoS$_2$*: V-MoS$_2$ was synthesized using a one-step CVD method. Liquid metal precursors were prepared by mixing two aqueous solutions containing Mo and V precursors, respectively (0.05 M sodium molybdate dihydrate (Na$_2$MoO$_4$·2H$_2$O), Sigma-Aldrich, 331058 and 0.05 M sodium orthovanadate dihydrate (Na$_3$VO$_4$·2H$_2$O), Sigma-Aldrich, S6508). These solutions were mixed in the given ratios to control the concentration of V in MoS$_2$ lattice. The mixed solution was spin-coated onto a SiO$_2$/Si wafer at 2500 revolution-per-minute for 1 min. For the growth of the V-MoS$_2$ film by CVD, the temperature was elevated to 850 °C under Ar atmosphere at a flow rate of 350 sccm and then dimethyl disulfide as a source of S and H$_2$ at flow rates of 3 and 5 sccm, respectively, were introduced for 10 min. After the growth of the V-MoS$_2$ film, the temperature was naturally cooled under Ar and H$_2$ atmosphere without changing the flow rate. The pristine MoS$_2$ film was synthesized by using the same CVD procedure without adding the V precursor.

*Transmission electron microscopy and specimen preparation*: Atomic-resolution ADF-STEM images of the samples were acquired using a probe aberration-corrected STEM (JEM-ARM200CF, Jeol Ltd.) operating at 80 keV. The detector angle range for ADF imaging was 45–180 mrad and the convergence semi-angle of the probe was 23 mrad. To avoid electron beam damage, the acquisition time of STEM image was conducted within 10 sec (Figure S27, Supporting Information). The multislice method was used for ADF-STEM image simulations, which is implemented in an open software, QSTEM software package[36] and the atomic quantifications from the ADFSTEM images were performed with commercial software qHAADF (HREM Research Ltd.). The TEM sample was prepared by the conventional transfer method with a poly (methyl methacrylate) (PMMA C4, MicroChem) supported layer.[18] After transferring the V-MoS$_2$ film onto the TEM grids (PELCO, 200 mesh, copper, 1.2 μm holes), the PMMA layer was removed by acetone. To avoid polymerization residuals during STEM imaging, the V-MoS$_2$ on the TEM grid was annealed at 300 °C for 3 h under the forming gas atmosphere prior to TEM analysis.

*Surface morphology and chemical state analysis:* The surface morphology of the as-grown V-MoS$_2$ film was examined by optical microscopy (Nikon LV-IM, Nikon) and scanning electron microscopy (JSM-7100F, JEOL). X-ray photoemission spectroscopy (K-Alpha, THERMO FISHER) was employed to characterize the elemental composition of V-doped MoS$_2$. Confocal

Raman and PL measurements were conducted using a Nanobase system (XperRam 100, Nanobase) with an excitation energy of 2.32 eV.

*Electrode fabrication:* The as-grown V-MoS$_2$ films (1 × 1 cm$^2$) were transferred onto the working electrode (graphite sheet) using a PMMA-supported wet-transfer method.[18] The PMMA was removed in hot acetone for 10 min to obtain a V-MoS$_2$/graphite sheet with a 1 × 1 cm$^2$ active geometric surface area. The procedure was repeated for transfer onto other electrodes (Cu and Ni).

*Electrochemical measurements:* All electrochemical measurements were conducted on a ZIVE SP2 (ZIVE Lab, Korea) electrochemical workstation within a three-electrode cell in 0.5 M H$_2$SO$_4$ at room temperature. A graphite substrate, saturated calomel electrode (SCE), and graphite rod were used as the working, reference, and counter electrodes, respectively. The electrolyte was de-aerated by purging with N$_2$ for 30 min prior to conducting the electrochemical experiment. Purging was maintained throughout the experiment. The catalytic behavior was characterized using LSV, EIS, CV, and chronoamperometry measurements. The LSV was measured in the range of 0 V to -0.8 V (*vs.* RHE) at a scan rate of 5 mV s$^{-1}$. EIS was measured from 1000 kHz to 10 mHz at an amplitude of 10 mV s$^{-1}$ and a constant potential of -0.3 V. A simple Randles circuit was applied to fit the EIS data using the software Zview. The C$_{dl}$ was measured by CV between 0.1 and 0.2 V (*vs.* RHE) at scan rates of 5, 10, 20, 30, 40, 50, 60, 80, and 100 mV s$^{-1}$. The stability and durability were studied using chronoamperometry at -0.25, -0.12, and -0.05 V and CV between -0.3 and 0.1 V (*vs.* RHE) at a scanning rate of 100 mV s$^{-1}$, respectively. All measured voltage values were converted to RHE with a value of 0.266 V using

$$E_{RHE} = E_{SCE} + E_{SCE}^0 + 0.0592 \times pH,$$

where $E_{RHE}$ is the converted potential value versus RHE, $E_{SCE}$ is the potential reading from the potentiostat, and $E^0_{SCE}$ is the experimentally determined standard electrode potential of SCE. A resistance test was conducted prior to measurements, and IR compensation was applied using ZIVE SP2 workstation software. The ohmic drop was corrected using the current interrupt method, and all potentials were IR- corrected with a compensation level of 90 %.

*Turnover-frequency (TOF) calculation:* According to a previous reference,[6] the hydrogen turnover-frequency (TOF) can be calculated using the formula:

$$\text{TOF}(s^{-1}) = \left( \frac{J(\text{A cm}^{-2})}{n \times N \times \text{relative EASA} \times (1.602 \times 10^{-19} C)} \right)$$

where n, N and relative EASA are the number of electrons required to evolve one mole of hydrogen molecule, the density of active sites, and the EASA of V-MoS$_2$ with respect to the Cu substrate, respectively.

$$\text{Relative EASA} = \frac{35.3\ mF\ cm^{-2}}{13.8\ mF\ cm^{-2}} = 2.73$$

We use the lattice parameters a = 3.192 Å and c = 13.378 Å for V-MoS$_2$, therefore the surface area of the unit cell is $5.78 \times 10^{-16}$ cm$^2$. Thus, the number of active sites were estimated to be about $1.73 \times 10^{15}$ cm$^{-2}$ assuming that the entire basal plane is catalytically active. Therefore, the geometric density of active sites of V-MoS$_2$ is

$$1.73 \times 10^{15}\ \text{cm}^{-2} \times 2.73 = 4.72 \times 10^{15}\ \text{cm}^{-2}$$

To estimate the TOF at the exchange current density, the TOF was extrapolated linearly from the TOF curve at 0 V.

*Computational methods:* The spin-polarized DFT calculations were conducted using the Vienna *ab initio* simulation package.[37] We employed the revised Perdew–Burke–Ernzerhof type exchange and correlation functional[38] combined with the introduction of vdW-DF[39] for the non-local correlation part, to accurately account for the dispersion interactions.[40] The projector augmented wave (PAW) method was used for ion interaction. The Brillouin zone was sampled using a 3 × 3 × 1 k-point mesh, while the electronic states were smeared using the Methfessel–Paxton scheme with a broadening width of 0.1 eV. A 6 × 6 × 1 k-point mesh was used for the DOS calculation. The electronic wave functions were expanded in a plane wave basis with a cutoff energy of 500 eV and the atomic relaxation was continued until the Hellmann–Feynman forces acting on the atoms were less than 0.02 eV Å$^{-1}$. 2D structures were modeled using a 5 × 5 supercell of MoS$_2$, and Cu and Ni substrates were modelled by employing slabs consisting of four atomic layers with 2D structures added. A 4 × 8 supercell of MoS$_2$ on top of a 5 × 10 supercell of the graphene model was used for the MoS$_2$/graphene structure. In each system, a vacuum layer of 16 Å was added onto the MoS$_2$ surface to eliminate possible interlayer interactions. The free energy of H adsorption, $\Delta G_{H^*}$, is defined as $\Delta G_{H^*} = \Delta E_H + \Delta E_{ZPE} - T\Delta S_H$, where $\Delta E_H$ is the H adsorption energy, $\Delta E_{ZPE}$ is the zero-point energy difference, T is room temperature, $\Delta S_H$ is the difference in entropy, and $\Delta E_{ZPE} - T\Delta S_H$ is 0.24 eV.[41] $\Delta E_H$ is defined as $\Delta E_H = E(H^* + \text{surface}) - E(\text{surface}) - 1/2 E(H_2)$, where E(H* + surface), E(surface), and E(H$_2$) represent the total energies of the H adsorbed surface, pristine surface, and H$_2$ molecule in the gas phase, respectively.

**Supporting Information**

Supporting Information is available from the author.

**Acknowledgements**

F. O. T. A. F. and S. J. Y. contributed equally to this work. K.K.K. acknowledges support from the Basic Science Research Program through the National Research Foundation of Korea (NRF), which is funded by the Ministry of Science, ICT & Future Planning (2018R1A2B2002302), and the Institute for Basic Science (IBS-R011-D1). Y.-M.K. acknowledges financial support by the Hydrogen Energy Innovation Technology Development Program of the National Research Foundation of Korea (NRF) funded by the Korean government (Ministry of Science and ICT (MSIT)) (No. 2019M3E6A1103959). Y.K.H. acknowledges support from the Basic Science Research Program through the National Research Foundation of Korea (NRF), which is funded by the Ministry of Science, ICT & Future Planning (2019R1A2C1008257). Y.H.L acknowledges support from the Institute for Basic Science (IBS-R011-D1). S.M.K. acknowledges support from the Basic Science Research Program through the National Research Foundation of Korea (NRF), which is funded by the Ministry of Science, ICT & Future Planning (2020R1A2B5B03002054).

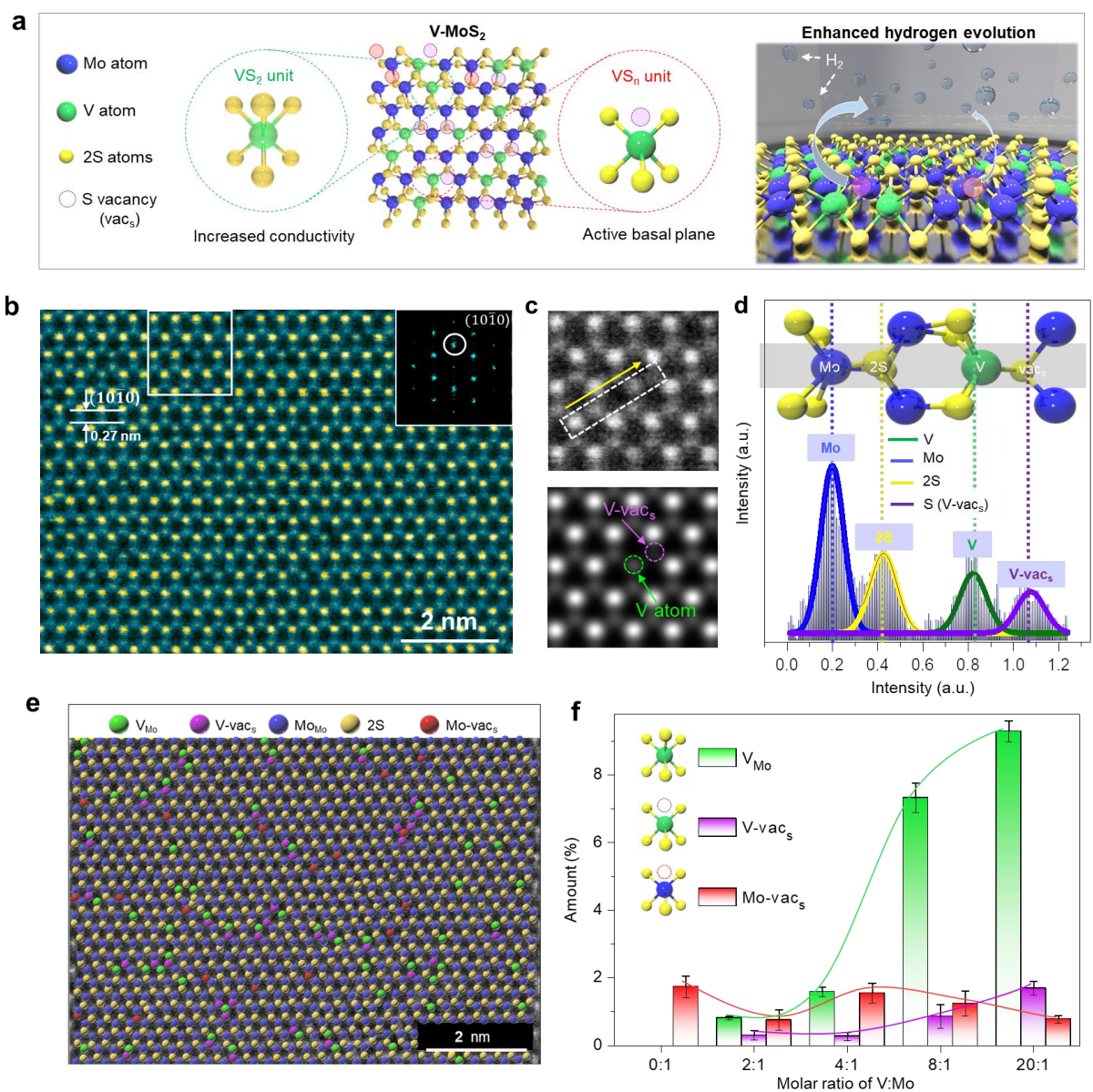

**Figure 1.** Atomic structure of monolayer V-MoS$_2$. a) Schematic of V-MoS$_2$ with VS$_2$ and VS$_n$ units and hydrogen evolution on V-MoS$_2$ via basal-plane activation. b) ADF-STEM image at 9.3% V concentration, indicating a d-spacing of 0.27 nm for 2H-MoS$_2$ and the corresponding electron-diffraction-pattern of $(10\bar{1}0)$ plane in the inset. c) STEM image of white square region in (b) and simulated image and d) the corresponding intensity profile. e) False-colored ADF-STEM image of monolayer V-MoS$_2$ with Mo-substituted V atom (V$_{Mo}$), sulfur-vacancy in V atom (V-vac$_s$), Mo atom (Mo$_{Mo}$), two S atoms (2S) and sulfur-vacancy in Mo atom (Mo-vac$_s$). f) Atomic % distribution of V$_{Mo}$, V-vac$_s$, and Mo-vac$_s$ as a function of molar ratio of V to Mo precursor. Statistical analysis data were obtained from false-colored ADF-STEM images.

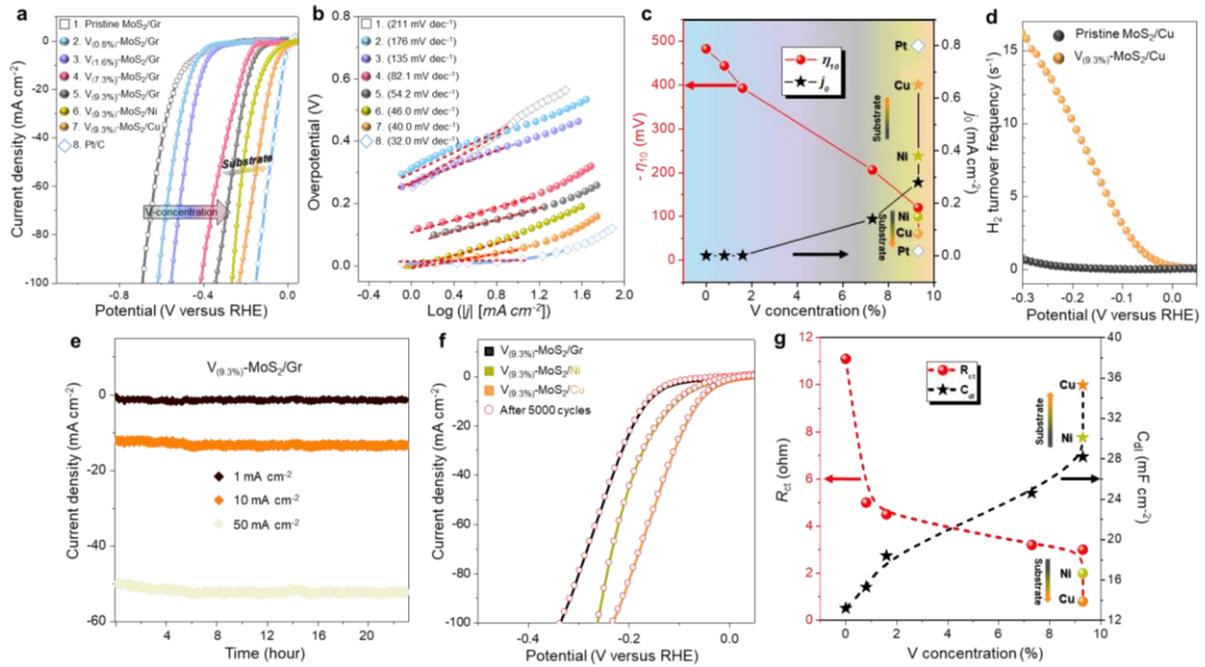

**Figure 2.** HER activity of V-MoS$_2$ in terms of V concentration and substrate. a) Polarization curves for pristine MoS$_2$ on graphite (Gr) substrate, V$_{(0.8\%)}$-MoS$_2$/Gr, V$_{(1.6\%)}$-MoS$_2$/Gr, V$_{(7.3\%)}$-MoS$_2$/Gr, V$_{(9.3\%)}$-MoS$_2$/Gr, V$_{(9.3\%)}$-MoS$_2$/Ni, V$_{(9.3\%)}$-MoS$_2$/Cu and Pt measured in N$_2$ saturated 0.5M H$_2$SO$_4$ electrolyte at 25°C at a scan rate of 5 mV s$^{-1}$. b) Tafel plots of V-MoS$_2$ extracted from polarization curves in (a). c) Overpotential at 10 mA cm$^{-2}$ ($\eta_{10}$) and exchange current density $j_0$ of V-MoS$_2$ samples. d) Turnover frequency of V$_{(9.3\%)}$-MoS$_2$/Cu compared to pristine MoS$_2$. e) Chronoamperometric curves of V$_{(9.3\%)}$-MoS$_2$/Gr under static current densities of 1, 10, and 50 mA cm$^{-2}$ for 24 h. f) Accelerated degradation test of V$_{(9.3\%)}$-MoS$_2$/Gr, Cu and Ni substrates after 5000 CV cycles. g) Charge transfer resistance ($R_{ct}$) and double layer capacitance ($C_{dl}$) of V-MoS$_2$ samples as a function of V concentration. The substrate dependence on Gr, Ni and Cu at V$_{(9.3\%)}$-MoS$_2$.

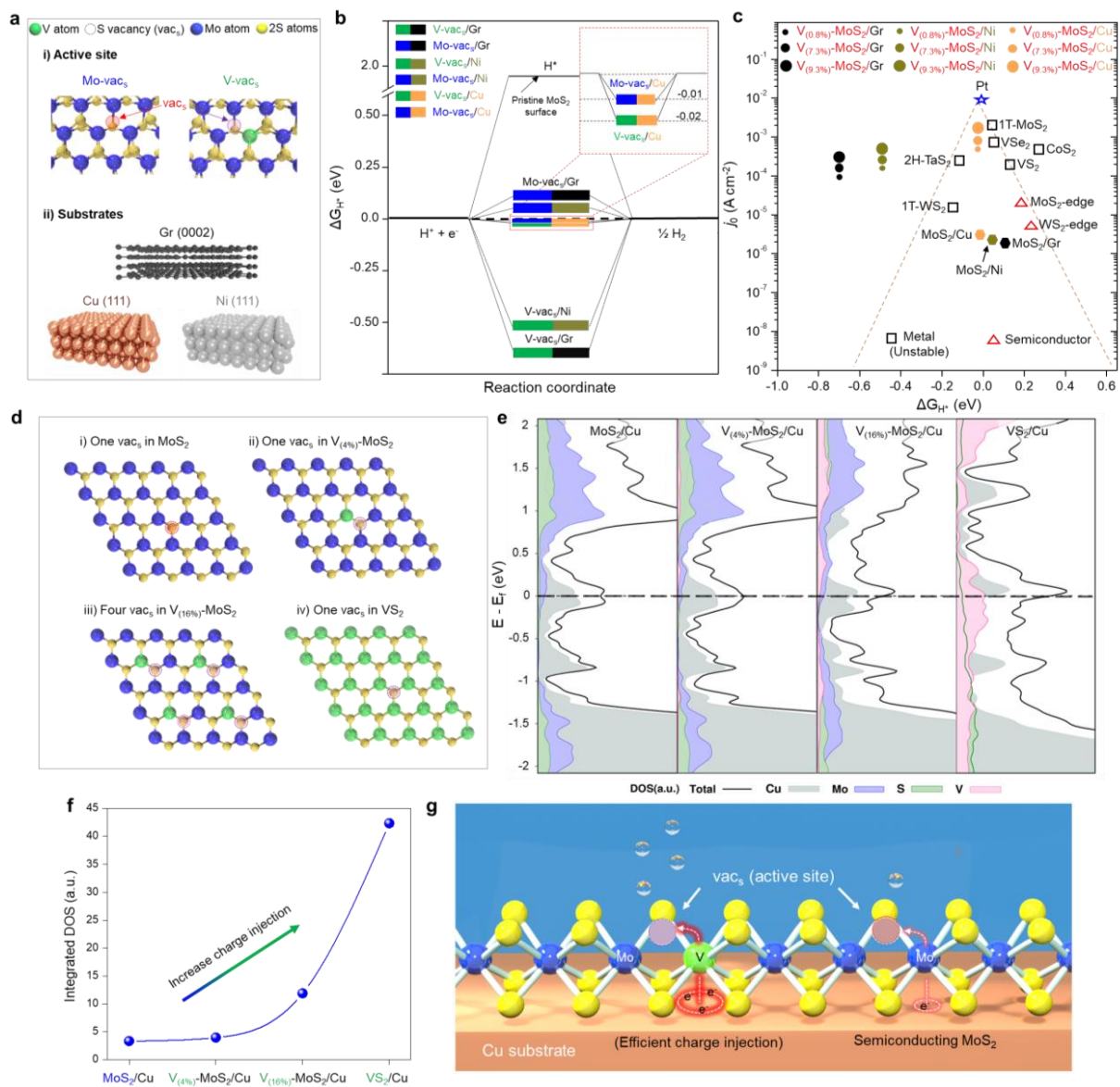

**Figure 3.** Gibbs free energy for hydrogen adsorption and density of states near the Fermi level of V-MoS$_2$ from DFT calculations. a) Schematic of active sites of sulfur vacancy in pristine MoS$_2$ and V-MoS$_2$ with substrates of Gr (0002), Cu (111), and Ni (111). b) Gibbs free energy diagram of V-MoS$_2$ on Ni, Cu, and Gr substrates. c) Volcano plot of other TMDs and our materials (V-MoS$_2$ on Cu, Ni, and Gr) with Gibbs free energy ($\Delta G_{H^*}$) and exchanged current density ($j_0$).[26,29-34] d) Supercell models for MoS$_2$, V-MoS$_2$ (low V and high V concentration), and VS$_2$ with sulfur vacancy. e) Projected density of states for Cu, Mo, S, and V atoms. f) Integrated density of states with V-MoS$_2$/Cu. g) Schematic of HER process in V-MoS$_2$/Cu catalyst.

**Table 1.** The comparison of catalytic parameters of V-MoS$_2$/Cu with Pt and other TMdCs. The comparison of Gibbs free energy for adsorbed hydrogen ($\Delta G_{H^*}$), overpotential at 10 mA cm$^{-2}$ current density ($\eta_{10}$), the exchange current density ($j_0$), charge transfer resistance ($R_{ct}$) and the turnover frequency (TOF) at 0 V of V-MoS$_2$, Pt and other TMdCs electrocatalyst.

| Catalyst | $\Delta G_{H^*}$ (eV) | $\eta_{10}$ (V) | $j_0$ (mA cm$^{-2}$) | $R_{ct}$ (Ω) | TOF @ 0 V (s$^{-1}$) | References |
|---|---|---|---|---|---|---|
| **Pt** | ~0.00 | -0.02 | 0.80 | 2.2 | 0.70 | Our work |
| **V-MoS$_2$/Cu** | -0.02 | -0.06 | 0.65 | 0.80 | 0.30 | Our work |
| **Nb$_{1.35}$S$_2$** | 0.11 | -0.15 | 0.80 | 7.40 | 0.20 | 6 |
| **1T WS$_2$** | 0.28 | -0.25 | 2.00 × 10$^{-3}$ | N/A | 0.04 | 31 |
| **VS$_2$** | -0.03 | -0.08 | N/A | 27 | N/A | 4 |
| **1T MoS$_2$** | 0.29 | -0.22 | 0.13 | 211 | N/A | 34 |

# Supporting Information

**Substitutional VS$_n$ Nanodispersed in MoS$_2$ Film for Pt-scalable Catalyst**

*Frederick Osei-Tutu Agyapong-Fordjour, Seok Joon Yun, Hyung-Jin Kim, Wooseon Choi, Soo Ho Choi, Laud Anim Adofo, Stephen Boandoh, Yong In Kim, Soo Min Kim, Young-Min Kim, Young Hee Lee\*, Young-Kyu Han\*, and Ki Kang Kim\**

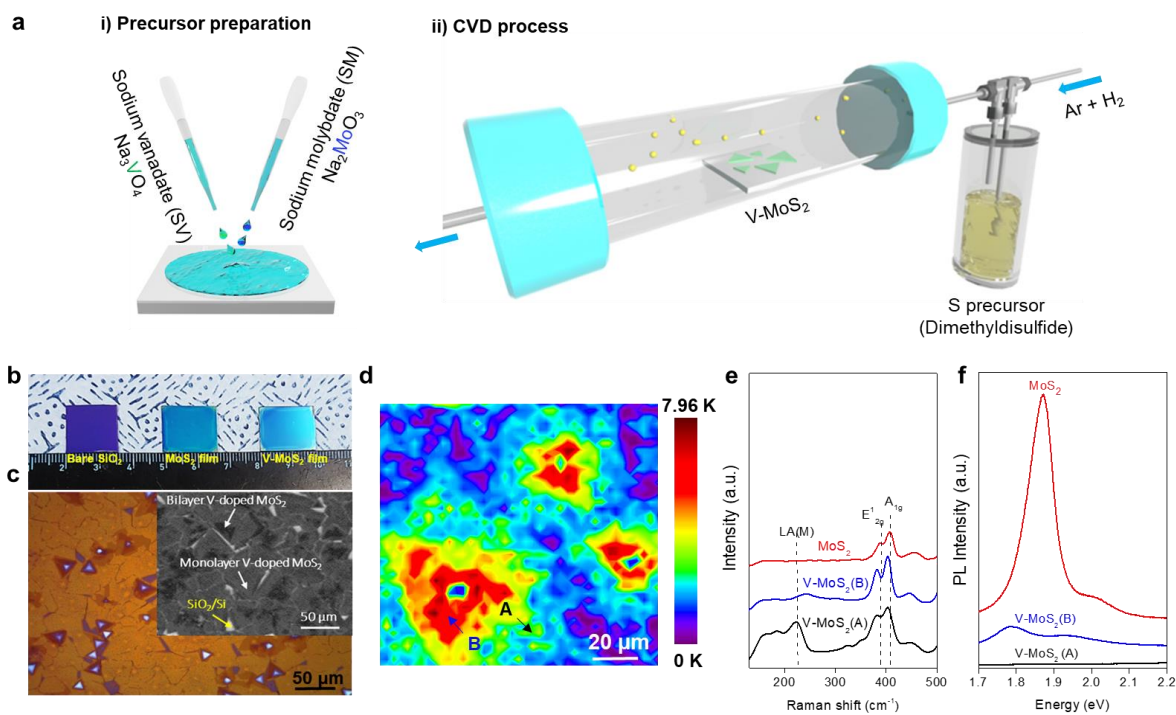

**Figure S1. CVD growth procedure and optical characterization of V-MoS$_2$.** a) Schematic of i) precursor preparation and ii) chemical vapor deposition for growing V-MoS$_2$ on SiO$_2$/Si substrate. b) Photograph of bare SiO$_2$/Si and an as-grown MoS$_2$ and V-MoS$_2$ film on SiO$_2$/Si substrate. c) Optical microscopy image of monolayer V-MoS$_2$ on SiO$_2$/Si substrate. Inset: SEM image of V-MoS$_2$ film showing V-MoS$_2$ film is predominantly monolayer with a small multilayer portion. d) Raman mapping image of V-MoS$_2$ for A$_{1g}$ mode (~398.4 cm$^{-1}$). Different colors in the mapping image indicate different thicknesses of V-MoS$_2$. e,f) Raman and PL spectra for pure MoS$_2$ and V-MoS$_2$ at positions A and B in d). The characteristic Raman peak of V-MoS$_2$ is newly developed at 225 cm$^{-1}$ and quenching of PL intensity as a result of the formation of metallic VS$_n$ units in MoS$_2$. [1]

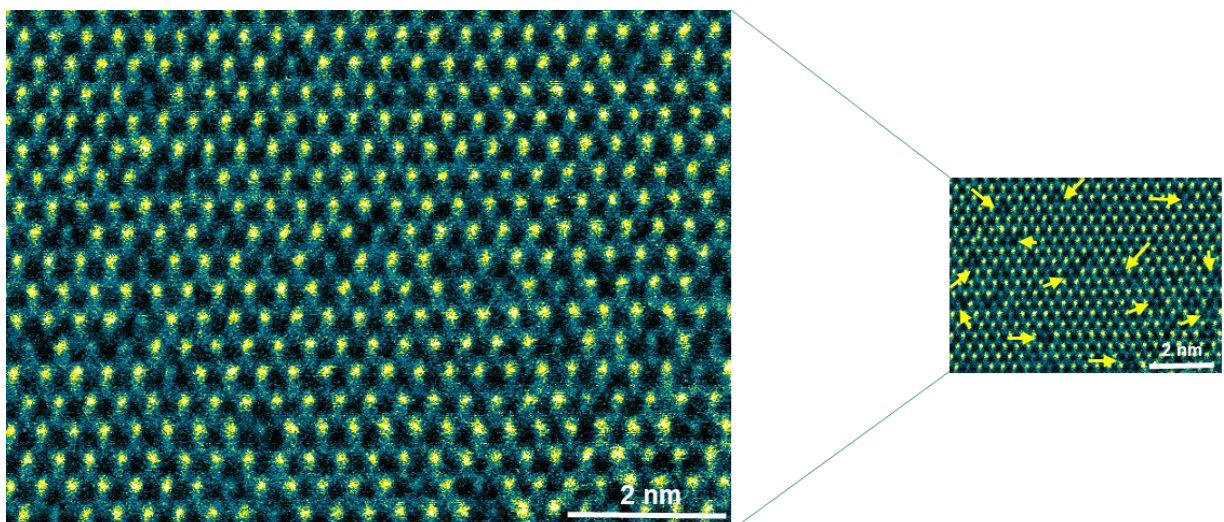

**Figure S2.** STEM image for $VS_n$ dispersion in V-$MoS_2$. The zoomed-out STEM image of V-$MoS_2$ clearly shows the nanodispersed $VS_n$ units indicated by yellow arrows.

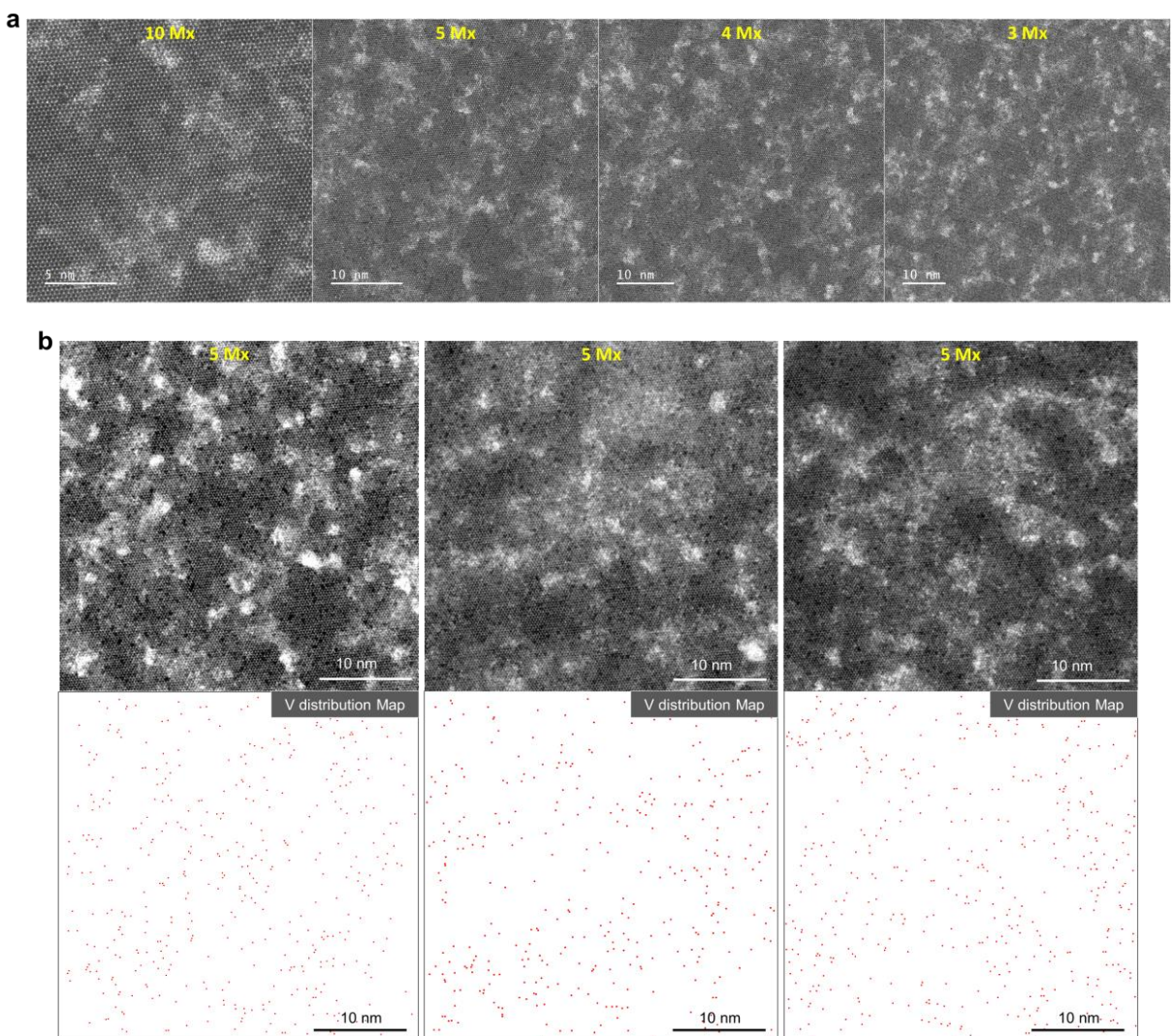

**Figure S3. Uniformity of VS$_n$ units in V-MoS$_2$ lattice**. a) STEM images of V$_{(9.3\%)}$-MoS$_2$ with various large field of views. The V-substitutions displayed as dark spots at Mo sites in V-MoS$_2$ are clearly observed in 10 Mx and 5 Mx images. Beyond 5 Mx could not identify V atoms. b) 5 Mx STEM images and corresponding V-distribution maps from three different regions. V atoms are homogenously distributed in MoS$_2$ lattice with scarce V aggregation formation. The white clusters on MoS$_2$ surface are carbon contaminants stemming from PMMA polymer which was used during the transfer process. [2]

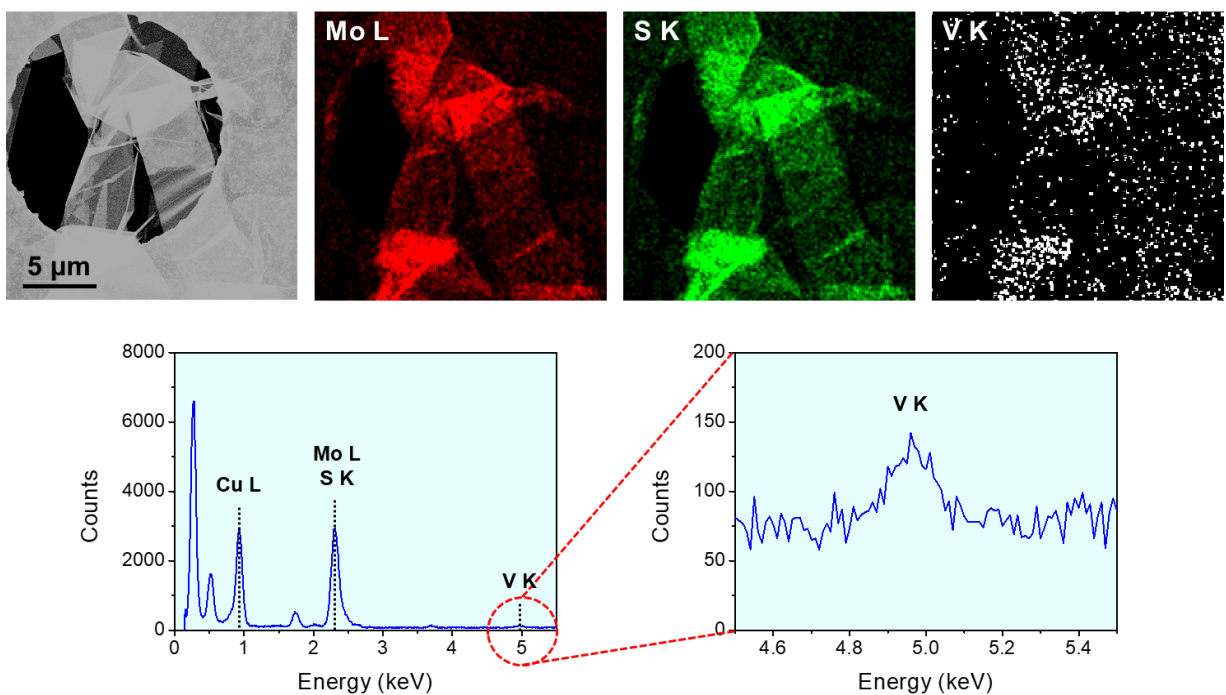

**Figure S4.** Energy dispersive X-ray spectroscopy analysis of V-MoS$_2$. The EDS mapping images and spectrum for V-MoS$_2$ prove the presences of Mo, S, and V atoms.

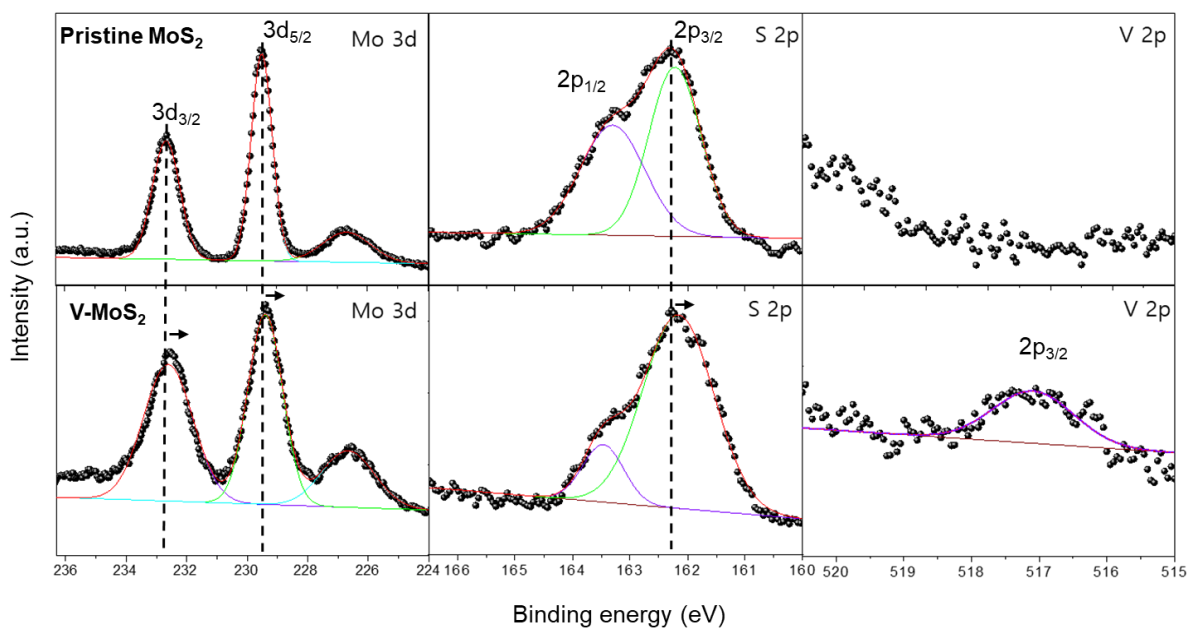

**Figure S5.** X-ray photoelectron spectroscopy (XPS) analysis of $MoS_2$ and V-$MoS_2$. XPS core-level spectra of Mo 3d, S 2p, and V 2p of $MoS_2$ and V-$MoS_2$, respectively. The two Mo 3d peaks and two S 2p peaks are $Mo^{4+}$ $3d_{5/2}$, $Mo^{4+}$ $3d_{3/2}$, S $2p_{3/2}$, and S $2p_{1/2}$ originating from $MoS_2$. The $V^{4+}$ $2p_{3/2}$ peak is clearly observed in V-$MoS_2$, implying the presence of V atoms in V-$MoS_2$. The slight red shift of the Mo 3d and S 2p peaks are attributed to the *p*-doping effect of V. [3,4]

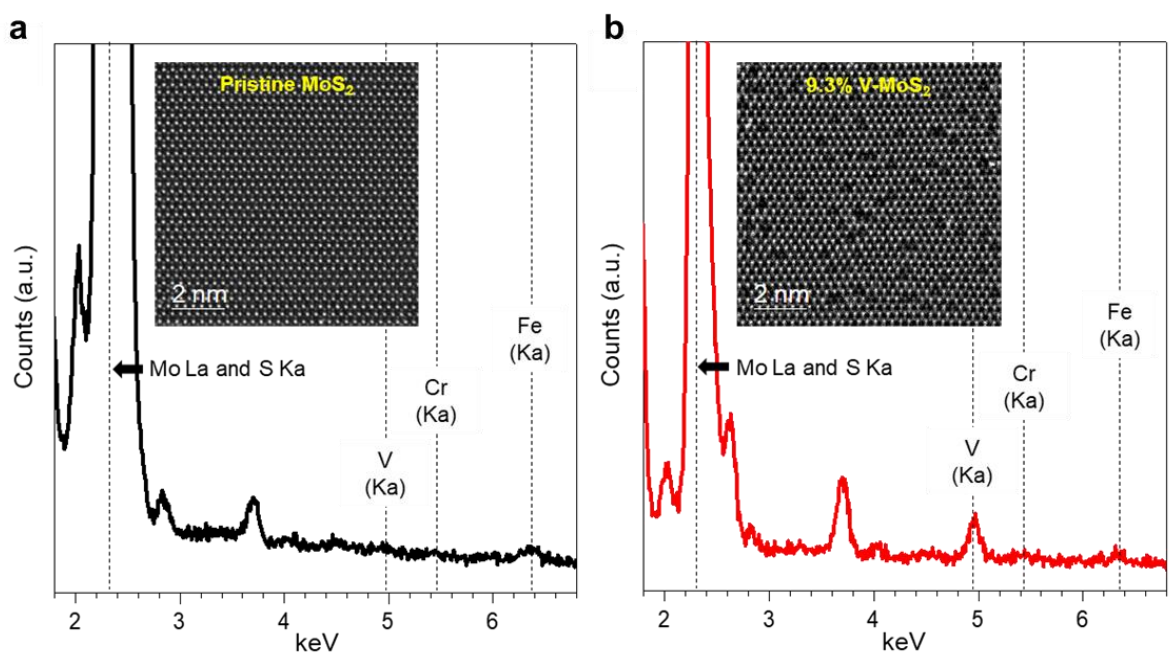

**Figure S6. Presence of V atoms in MoS$_2$ lattice.** EDX spectra for a) pristine and b) V-MoS$_2$ with representative STEM images. While dark spots (V atoms) in STEM image of pristine MoS$_2$ are rarely found, the distinct spots are observed in V-MoS$_2$. Furthermore, the prominent V Ka peak is clearly visible in the EDX spectrum of V-MoS$_2$. The tiny Cr and Fe Ka peaks detected in both EDX spectra probably come from metal parts of objective lens and EDX holder (known as instrumental peaks),[5] rather than impurities in the MoS$_2$ lattice.[6,7] Therefore, the dark spots are V atoms, not Cr or Fe atoms.

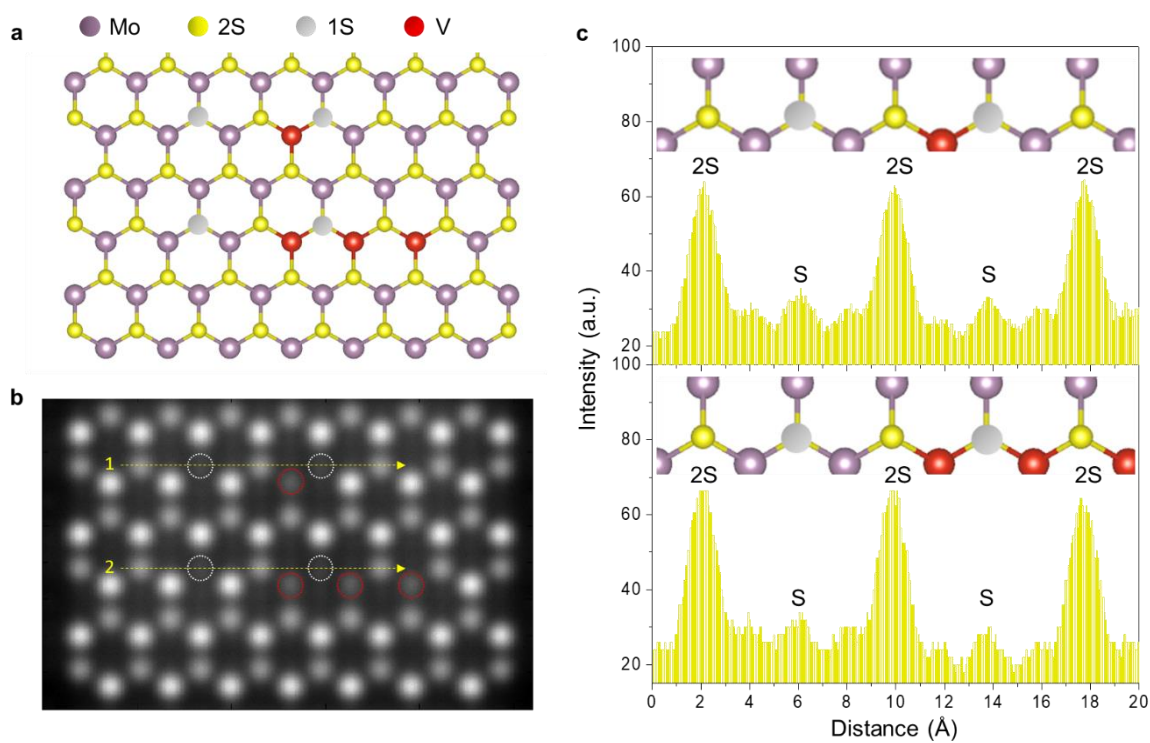

**Figure S7. Intensity profile of 1S and 2S next to Mo and V atoms.** a) Schematic ball-and-stick model of V-MoS$_2$ with Vac$_s$ and b) corresponding simulated image. c) Intensity profiles of 1S and 2S atoms along lines 1 and 2 in b. Even though 1S and 2S next to V atom are slightly dimmer compared to those next Mo, they are still clearly distinguished by the intensity.

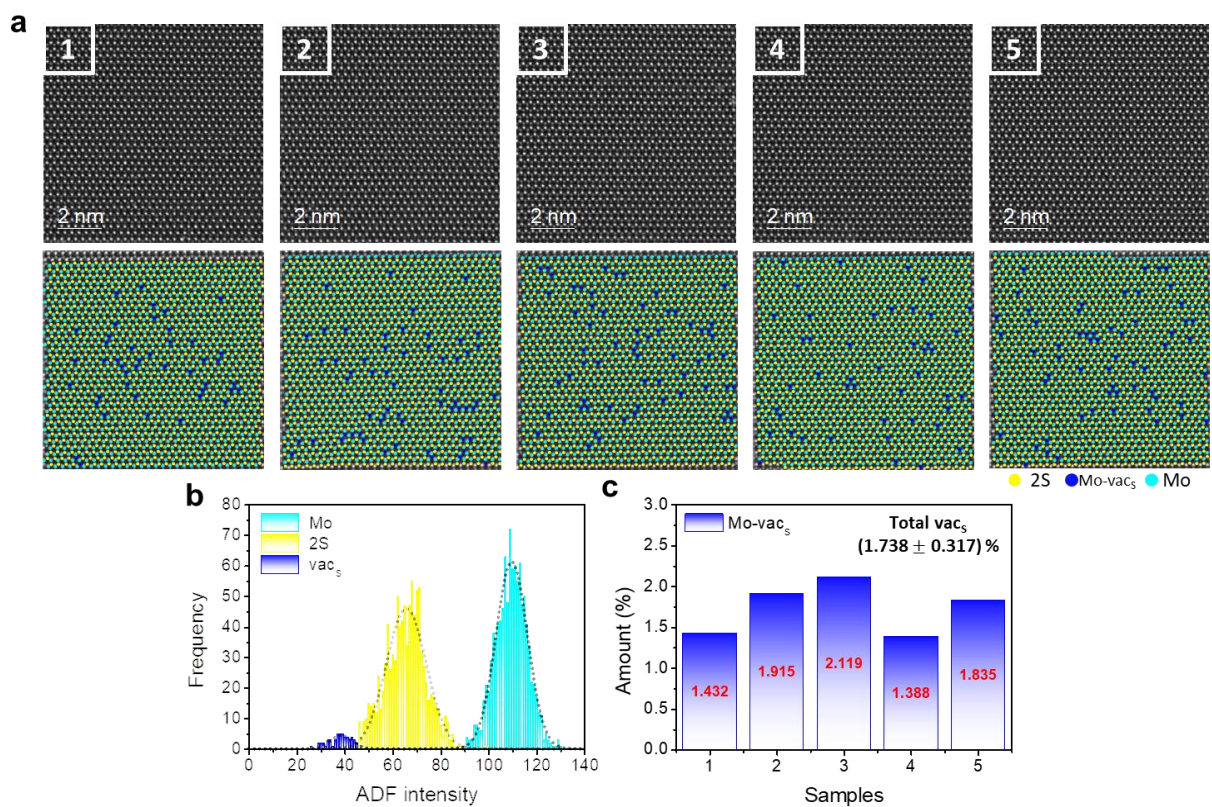

**Figure S8. Atom site-wise mapping of pristine MoS$_2$ monolayer.** a) Weiner-filtered ADF-STEM images (top) and false-colored STEM images (bottom) of MoS$_2$. b) Frequency of Mo, S, and vac$_s$ sites with their ADF intensities. c) Atomic % of sulfur vacancies in pristine MoS$_2$ (Mo-vac$_s$).

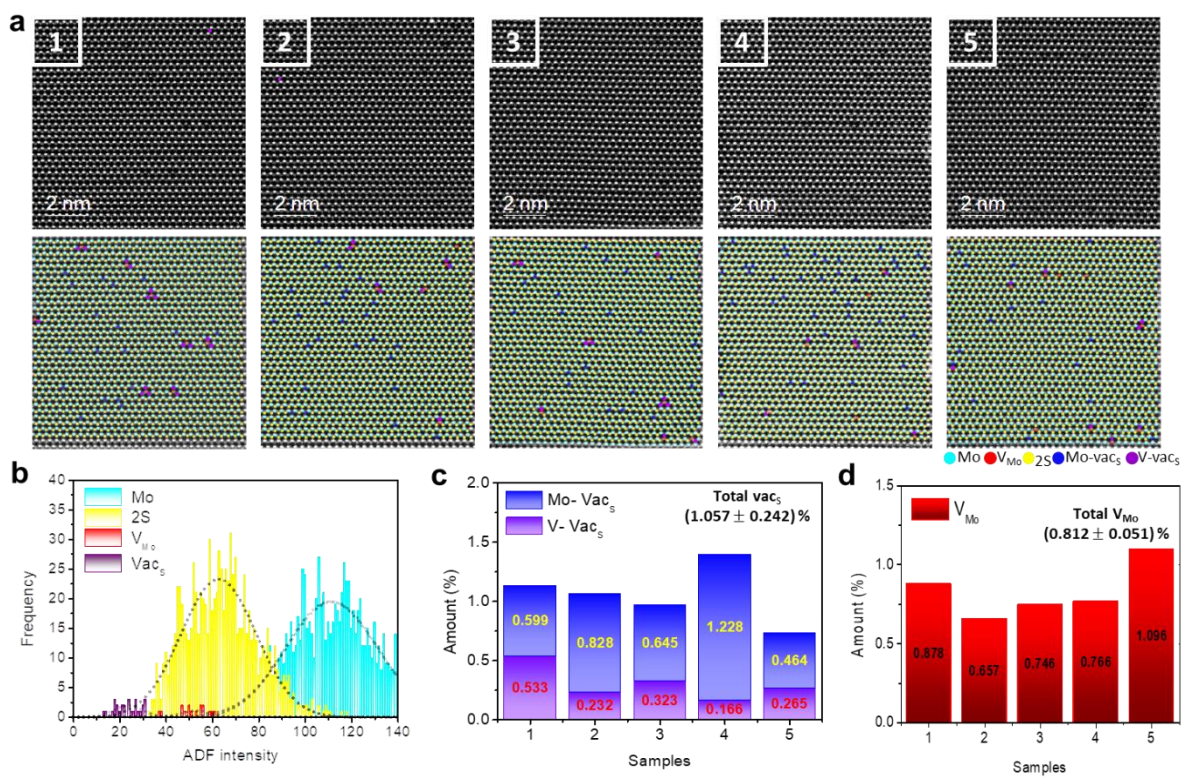

**Figure S9. Atom site-wise mapping of 2:1 V-MoS$_2$ (V:Mo)**. a) Weiner-filtered ADF-STEM images (top) and false-colored STEM images (bottom) of 2:1 V-MoS$_2$. b) Frequency of Mo, S, V$_{Mo}$, and vac$_s$ sites with their ADF intensities. c,d) Atomic % of Mo-vac$_s$ and V-vac$_s$ and V$_{Mo}$, respectively.

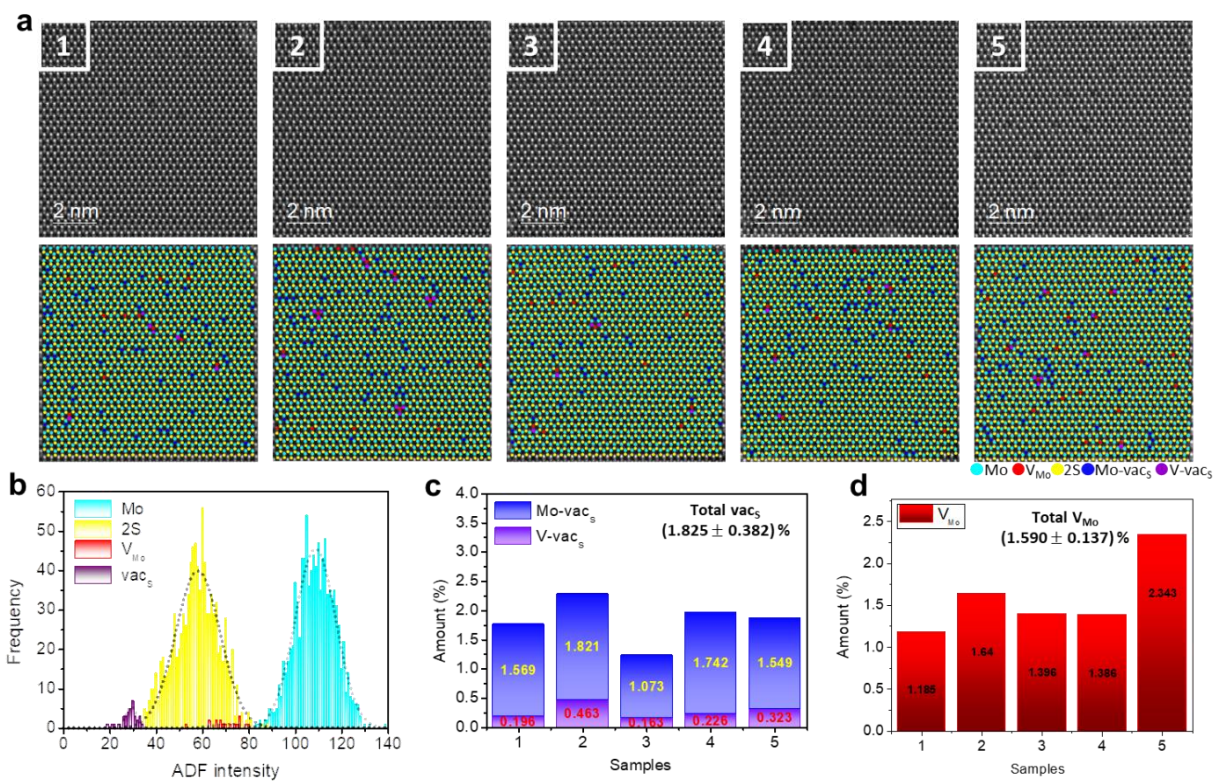

**Figure S10. Atom site-wise mapping of 4:1 V-MoS$_2$ (V:Mo)**. a) Weiner-filtered ADF-STEM images (top) and false-colored STEM images (bottom) of 4:1 V-MoS$_2$. b) Frequency of Mo, S, V$_{Mo}$, and vac$_s$ sites with their ADF intensities. c,d) Atomic % of Mo-vac$_s$ and V-vac$_s$ and V$_{Mo}$, respectively.

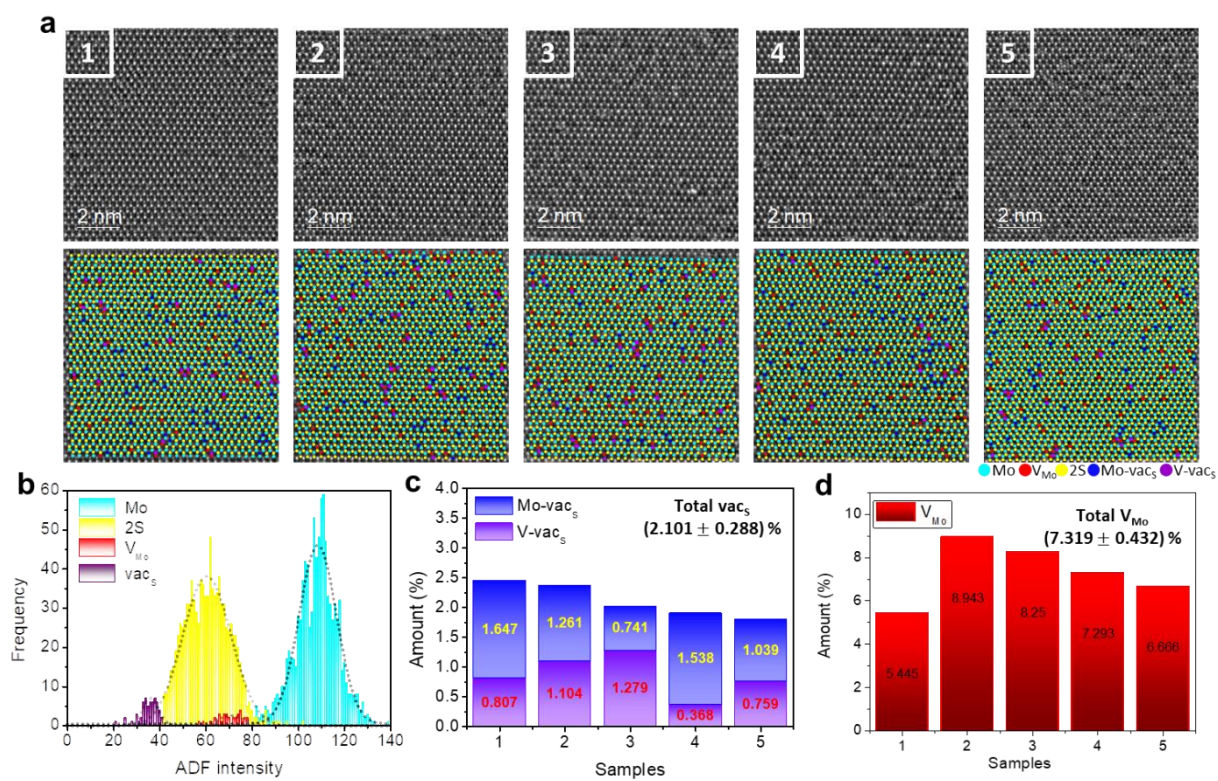

**Figure S11. Atom site-wise mapping of 8:1 V-MoS$_2$ (V:Mo)**. a) Weiner-filtered ADF-STEM images (top) and false-colored STEM images (bottom) of 8:1 V-MoS$_2$. b) Frequency of Mo, S, V$_{Mo}$ and vac$_s$ sites with their ADF intensities. c,d) Atomic % of Mo-vac$_s$ and V-vac$_s$ and V$_{Mo}$, respectively.

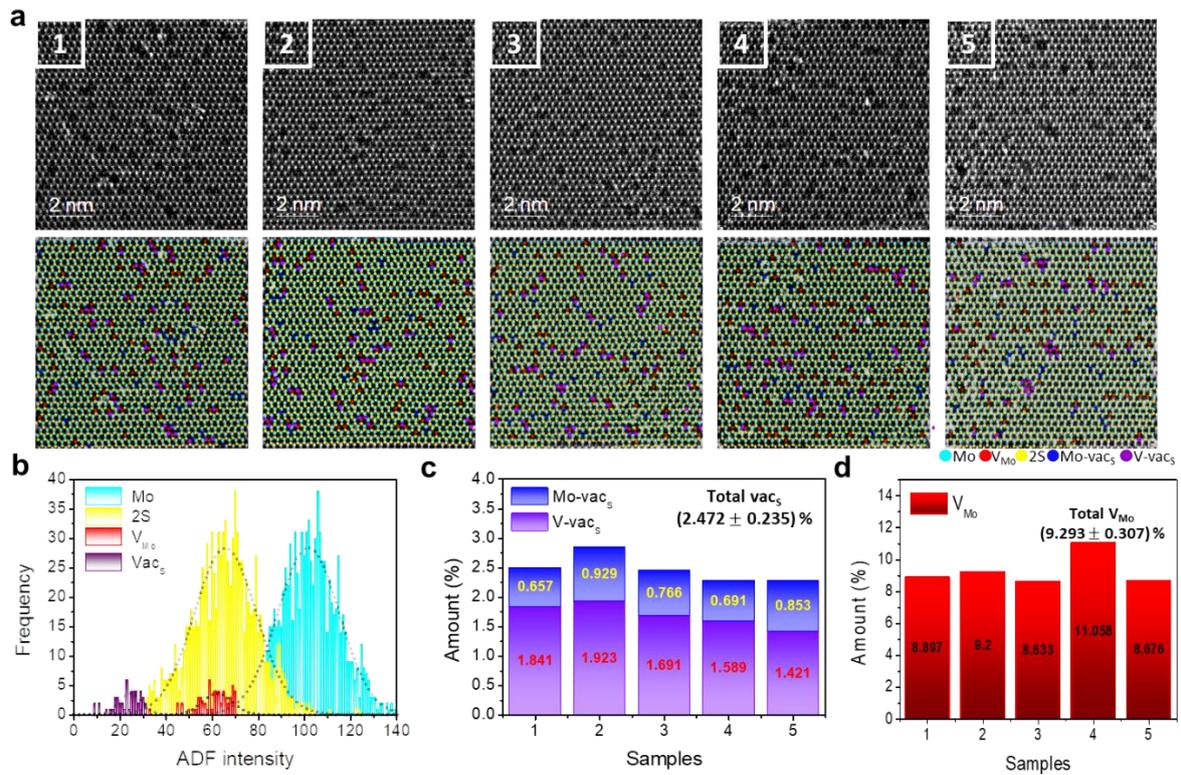

**Figure S12. Atom site-wise mapping of 20:1 V-MoS$_2$ (V:Mo)**. a) Weiner-filtered ADF-STEM images (top) and false-colored STEM images (bottom) of 20:1 V-MoS$_2$. b) Frequency of Mo, S, V$_{Mo}$, and vac$_s$ sites with their ADF intensities. c,d) Atomic % of Mo-vac$_s$ and V-vac$_s$ and V$_{Mo}$, respectively.

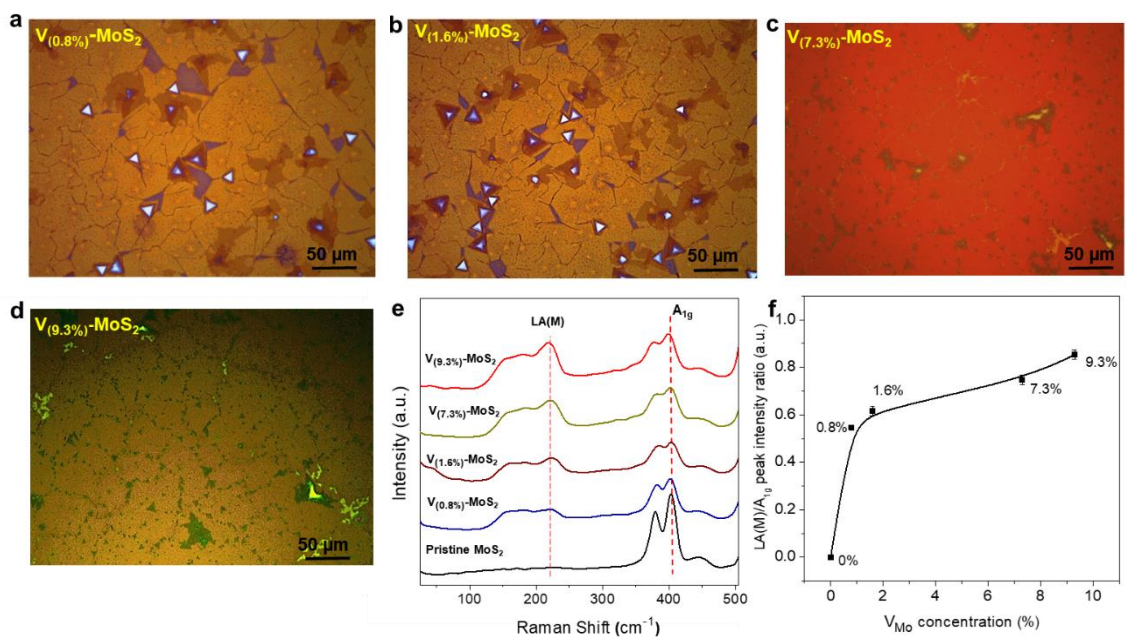

**Figure S13. Optical microscopy image and Raman spectra of V-MoS$_2$ synthesized with different V** at%. a-d) Optical microscopy images of as-grown V-MoS$_2$ film with 0.8, 1.6, 7.3, and 9.3 V at%, respectively. e) Raman spectra of as-grown V-MoS$_2$ film with various V:Mo mixing ratio. f) Peak intensity ratio of LA(M) mode (~225 cm$^{-1}$) to A$_{1g}$ as a function of the V concentration.[1]

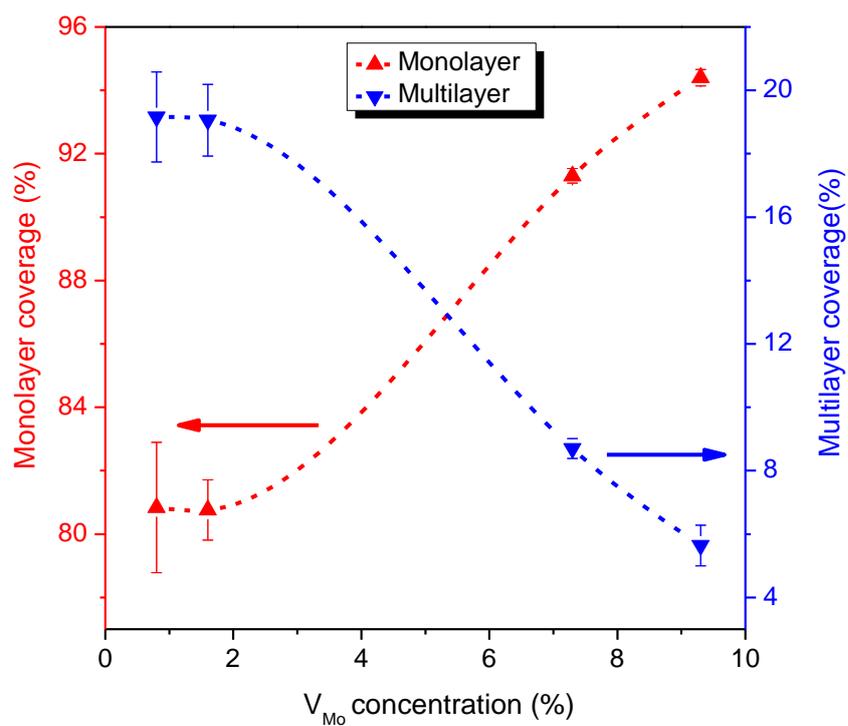

**Figure S14.** Statistics of monolayer and multilayer extracted from optical images using commercial Gwyddion software.[8]

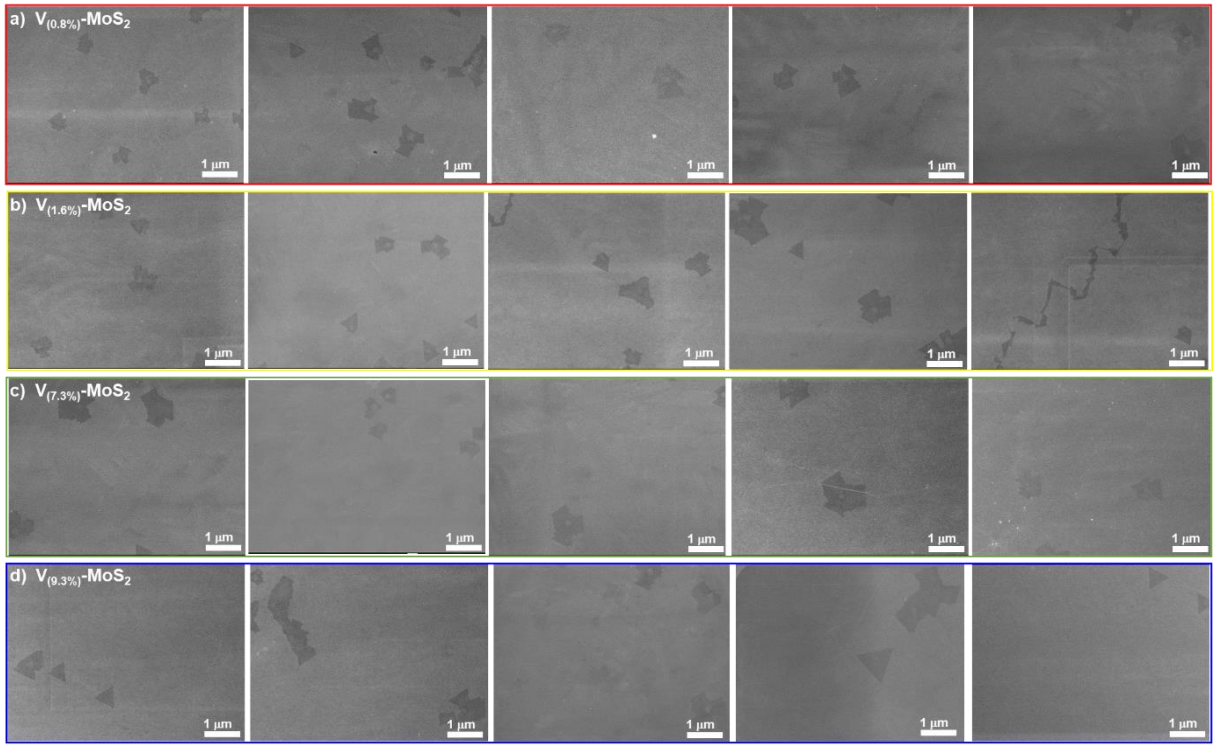

**Figure S15. Observation of multilayer (dark) regions in V-MoS$_2$.** a-d) SEM images of as-grown V-MoS$_2$ film with different V concentrations of 0.8, 1.6, 7.3, and 9.3 V at%.

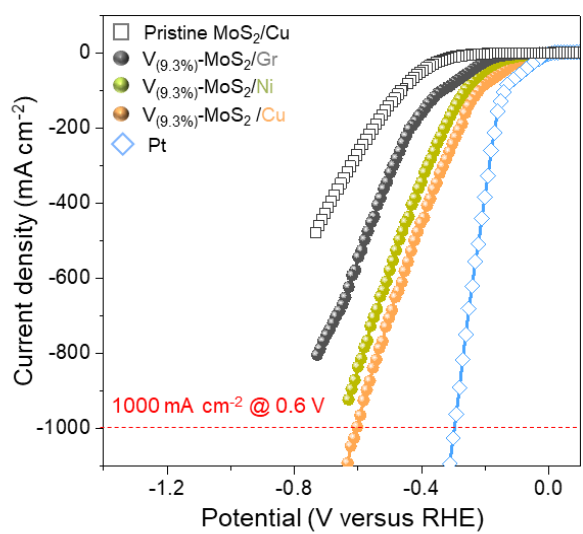

**Figure S16. High current density of V-MoS₂ catalyst on various substrate**. The high current density exceeding 1,000 mA cm$^{-2}$ at 0.6 V for V$_{(9.3\%)}$-MoS$_2$/Cu demonstrates its suitability for industrial application.

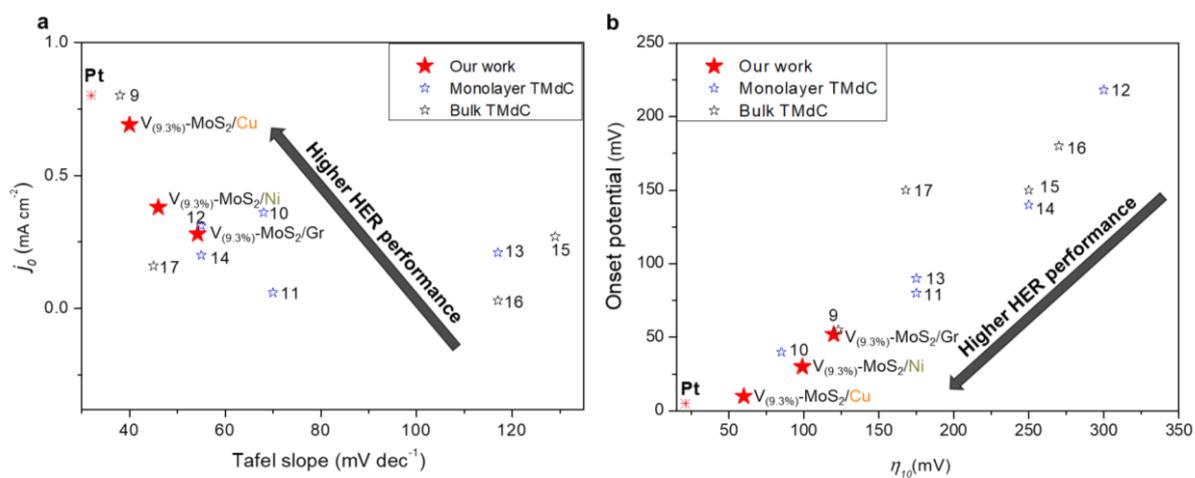

**Figure S17. Comparison of electrocatalytic parameters of V-MoS$_2$ film with those of other 2D TMdCs**. a) Exchange current density ($j_0$) against Tafel slope. b) Onset potential versus overpotential for 10 mA cm$^{-2}$ current density ($\eta_{10}$) of our materials compared with other 2D electrocatalysts. V$_{(9.3\%)}$-MoS$_2$ film on the Cu substrate exhibits extremely low onset potential of -10 mV, similar to that of Pt (-5 mV). Comparative data were collected from the literatures.[9-17]

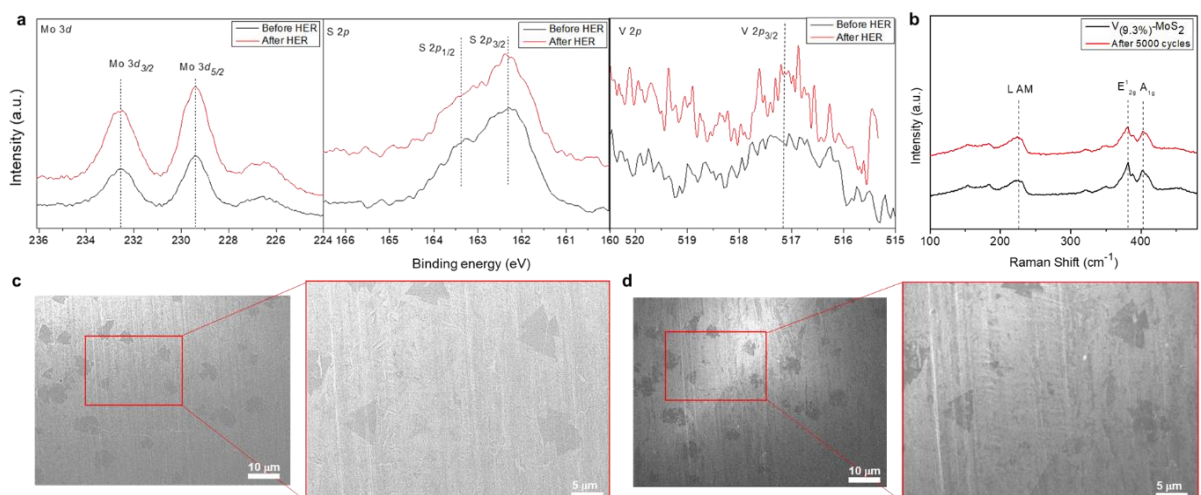

**Figure S18. Stability of V$_{(9.3\%)}$-MoS$_2$ before and after HER cycling test.** a) XPS core-level spectra of V$_{(9.3\%)}$-MoS$_2$ before and after HER cycling-test for Mo 3d, S 2p, and V 2p, respectively. b) Representative Raman spectra analysis of V$_{(9.3\%)}$-MoS$_2$ before and after 5000 HER cycling test. c,d) SEM image of V-MoS$_2$ c) before and d) after 5000 cycling with corresponding zoomed-in images. No significant changes are observed in the XPS core-level spectra, Raman spectra, and SEM images after HER cycle test. These indicate that V-MoS$_2$ is stable during HER.

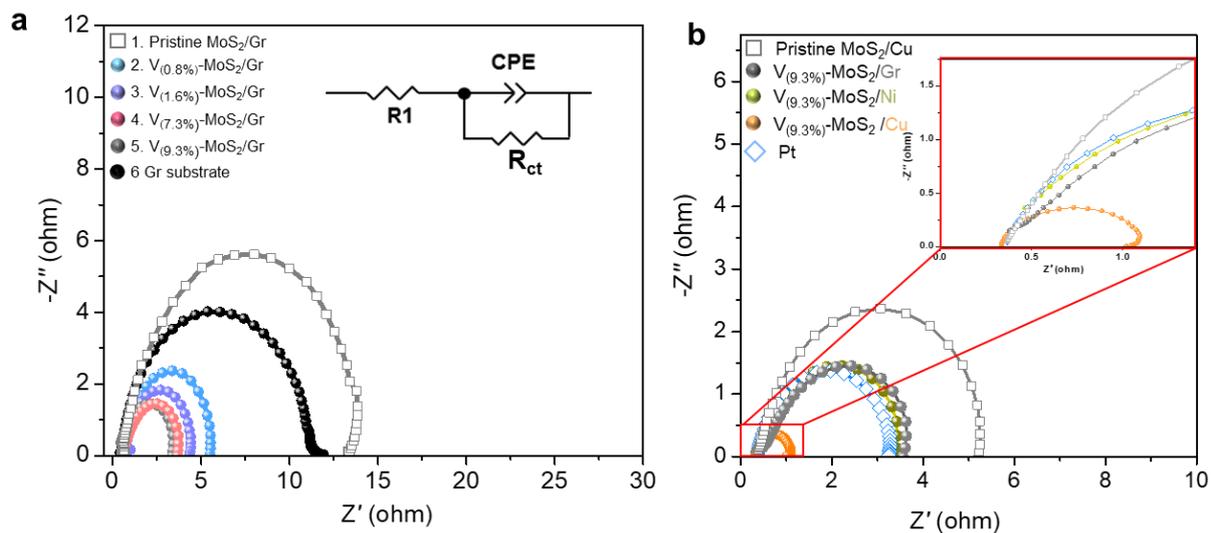

**Figure S19. Charge transfer resistance analysis of V-MoS$_2$ in terms of the V concentration and the substrate**. a,b) Nyquist plots for V-MoS$_2$/Gr and V$_{(9.3\%)}$-MoS$_2$ on Gr, Ni, and Cu compared to Pt. Inset: The equivalent Randel circuit model is used to extract the series (R1) and charge transfer resistance (R$_{ct}$).

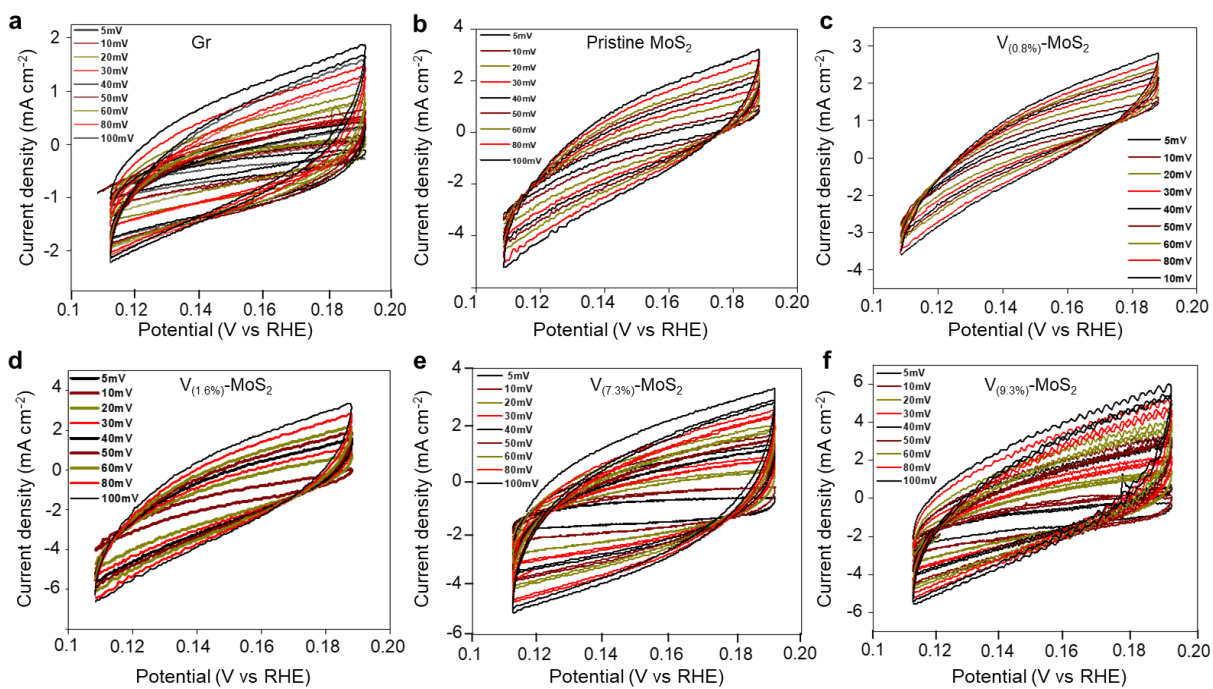

**Figure S20. Cyclic voltammetry measurement for double-layer capacitance.** a-f) Cyclic voltammograms of Gr, pristine $MoS_2$, $V_{(0.8\%)}$-$MoS_2$, $V_{(1.6\%)}$-$MoS_2$, $V_{(7.3\%)}$-$MoS_2$, and $V_{(9.3\%)}$-$MoS_2$ on Gr.

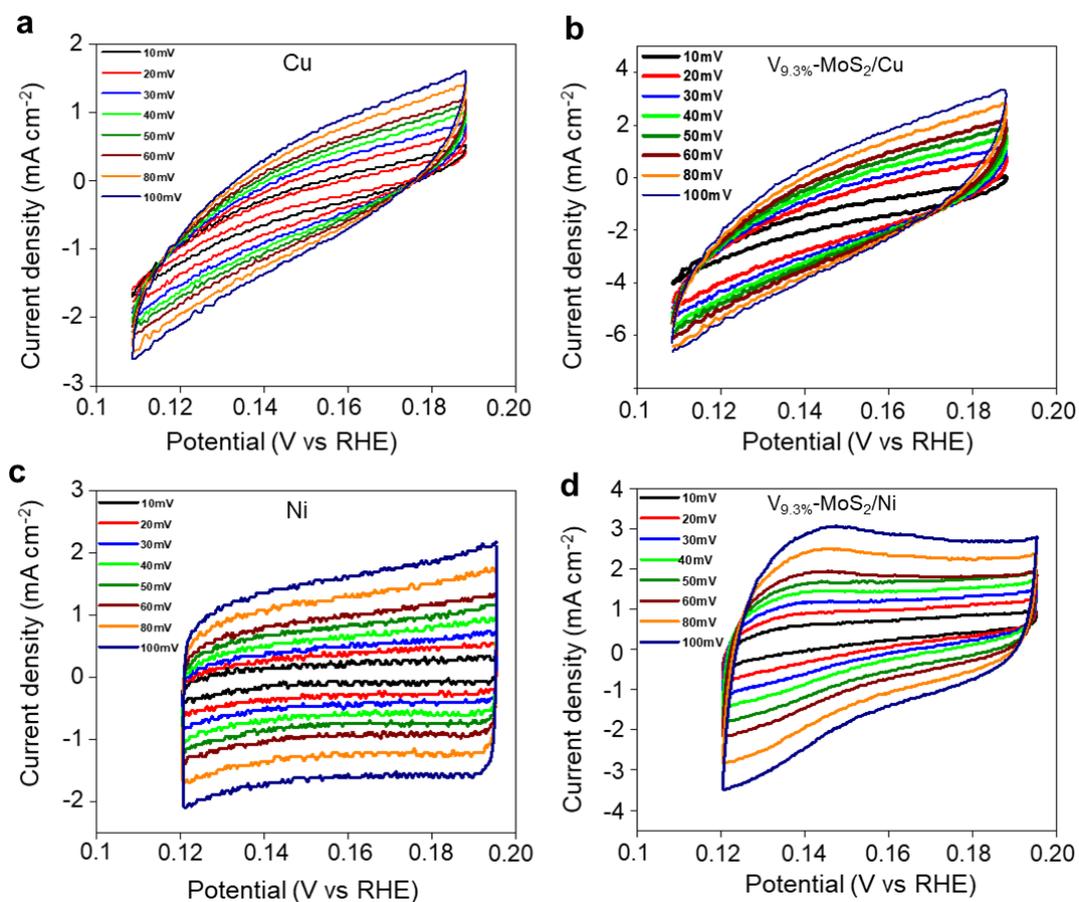

**Figure S21. Cyclic voltammetry measurement for double-layer capacitance.** a-d) Cyclic voltammograms of bare Cu, V$_{(9.3\%)}$-MoS$_2$/Cu, bare Ni, and V$_{(9.3\%)}$-MoS$_2$/Ni, respectively.

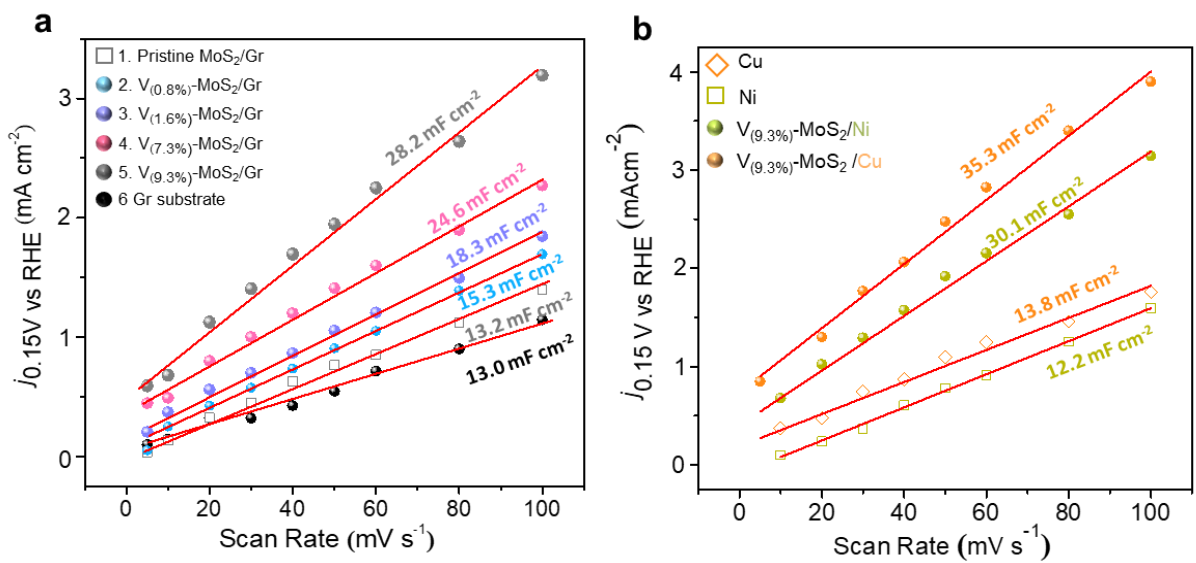

**Figure S22. Double-layer capacitance of V-MoS$_2$ with respect to the V concentration and the substrates.** a,b) Capacitive current of V-MoS$_2$ on Gr substrate for various V concentrations and V$_{(9.3\%)}$-MoS$_2$ on Cu and Ni substrates as a function of the scan rate. The capacitive current was extracted at 0.15 V vs RHE in Figures S20 and S21, Supporting Information.

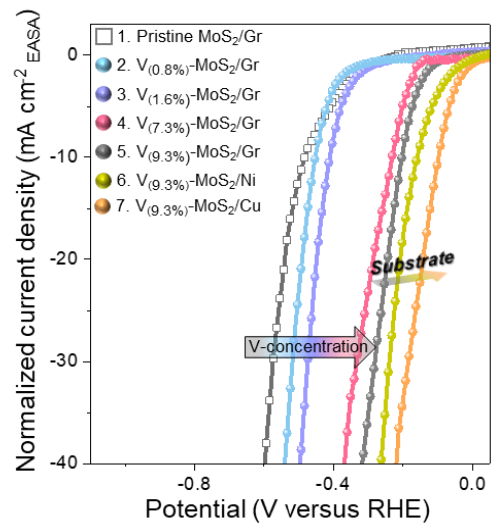

**Figure S23.** Relative EASA normalized polarization curves for pristine MoS$_2$ on graphite (Gr) substrate, V$_{(0.8\%)}$-MoS$_2$/Gr, V$_{(1.6\%)}$-MoS$_2$/Gr, V$_{(7.3\%)}$-MoS$_2$/Gr, V$_{(9.3\%)}$-MoS$_2$/Gr, V$_{(9.3\%)}$-MoS$_2$/Ni, V$_{(9.3\%)}$-MoS$_2$/Cu.

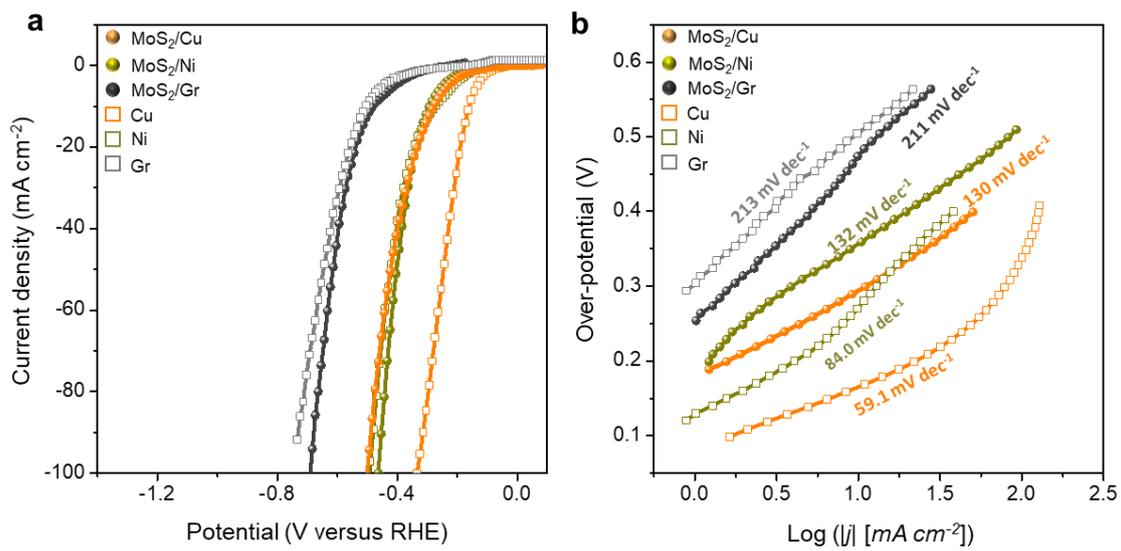

**Figure S24.** a,b) Polarization and Tafel curves for pristine MoS$_2$ on Gr, Cu, and Ni substrates.

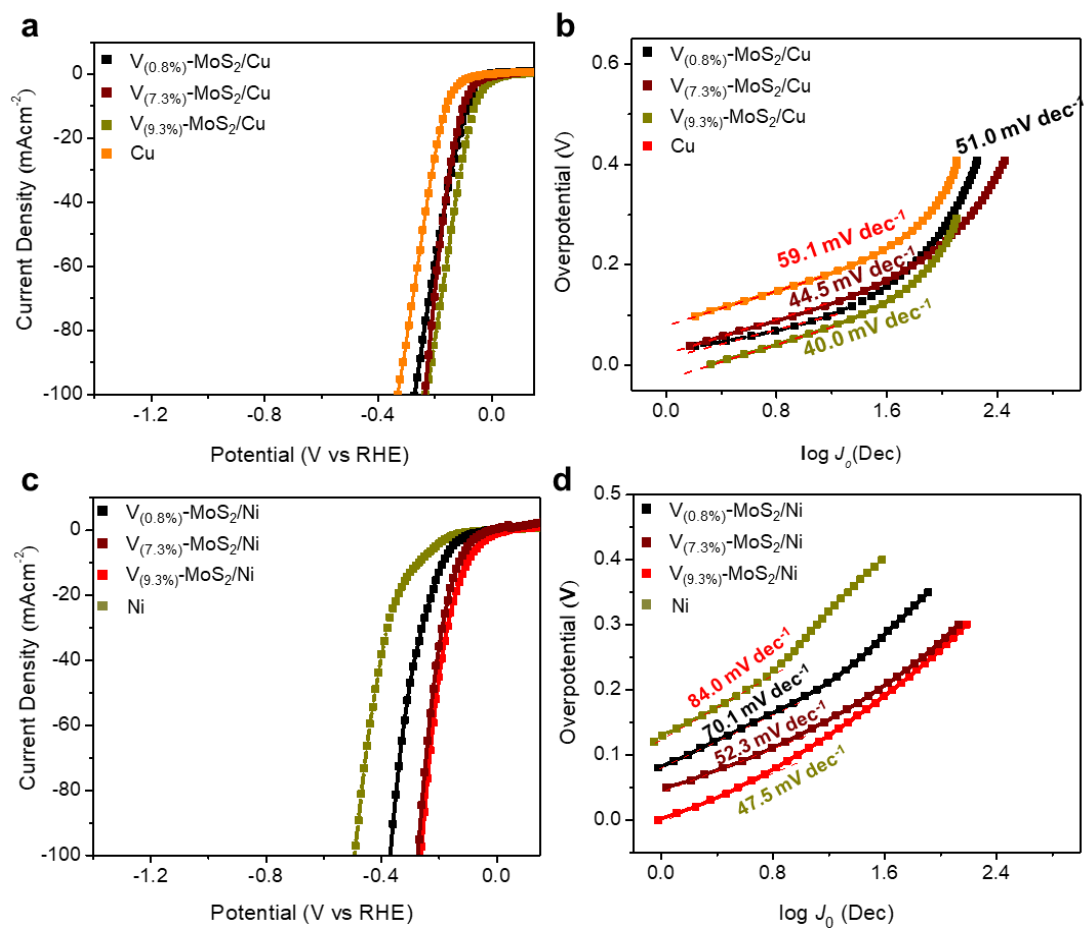

**Figure S25. Polarization and Tafel curves of V-MoS₂ on Cu and Ni substrates**. a,c) Polarization and b,d) Tafel curves for $V_{(0.8\%)}$-MoS$_2$, $V_{(7.3\%)}$-MoS$_2$, and $V_{(9.3\%)}$-MoS$_2$ on Cu and Ni substrates, respectively.

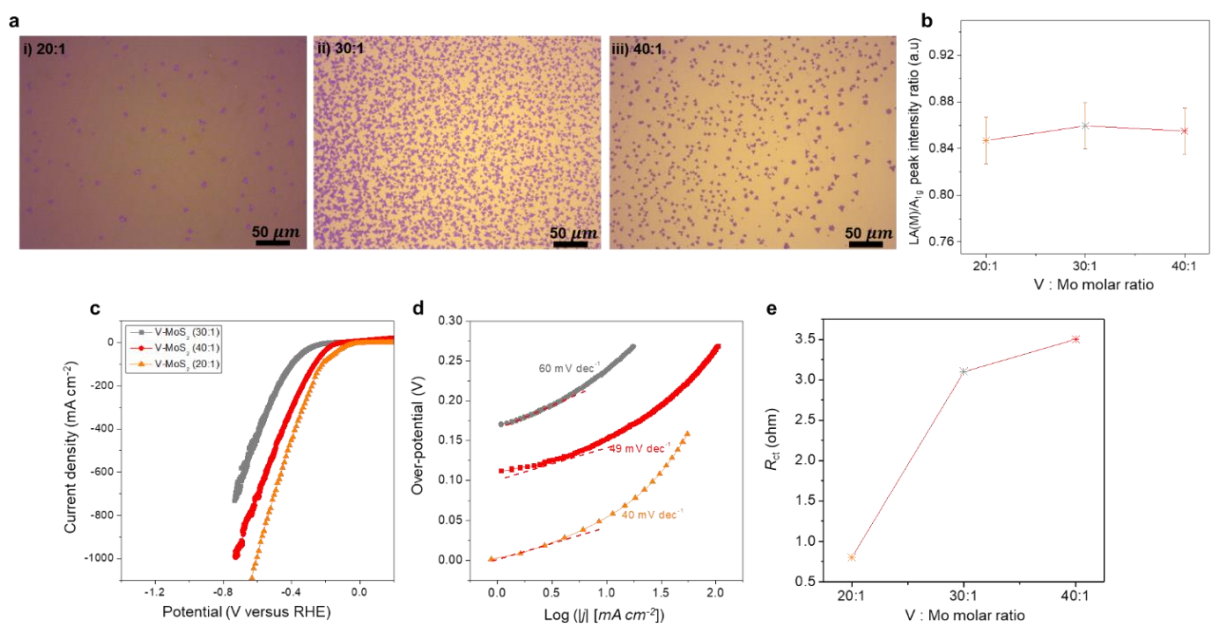

**Figure S26. HER performance with higher V to Mo precursor molar ratio**. a) Optical microscopy images of as-grown V-MoS$_2$ with higher molar ratio of V to Mo precursor of 20:1, 30:1 and 40:1, respectively. While the V-MoS$_2$ film was grown with the molar ratio of 20:1, only V-MoS$_2$ flakes were achieved with higher molar ratio of vanadium precursor. b) Raman intensity ratio of LA(M)/A$_{1g}$ for as-grown V-MoS$_2$. c-e) Polarization and Tafel curves and charge transfer resistance of V-MoS$_2$ on Cu substrate.

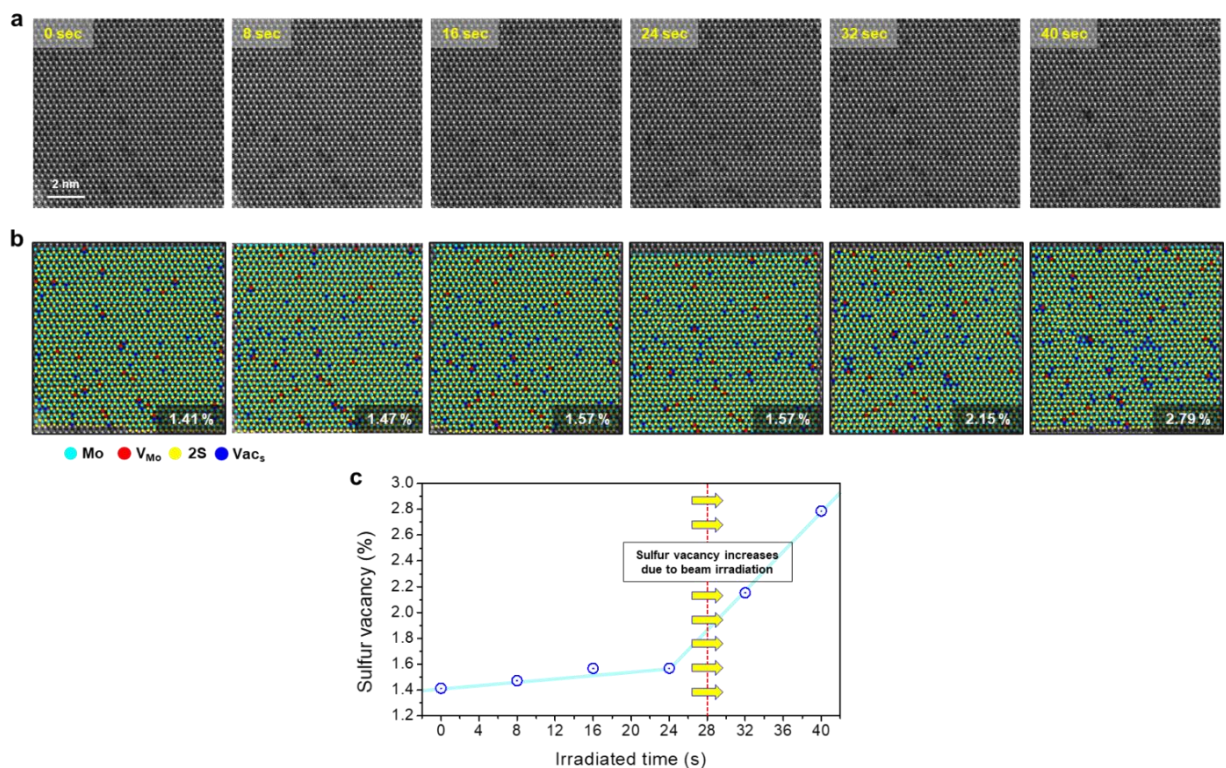

**Figure S27. Time-dependent e-beam irradiation analysis using continuous electron beam scanning at 80 kV with a probe current of 25 pA.** a) Continuously acquired ADF-STEM images and b) corresponding atomic maps. c) Statistics for S vacancy concentration with irradiation time in sec. In this work, ADF-STEM was acquired within 10 sec with a scanning rate of 8 μs/pixel for a 1024 × 1024 pixels resolution. Therefore, electron beam damage of the sample can be excluded.

**Table S1.** S vacancy (vac$_s$) formation energy of MoS$_2$ with secondary cation relative to that of pristine MoS$_2$. The value of 0.22 eV for V means that the formation of vac$_s$ next to a V atom is more stable than next to Mo atom.

| Cation | ΔE (eV) |
|--------|---------|
| V      | 0.22    |
| Nb     | 0.12    |
| Ta     | 0.05    |

**Table S2.** Comparison of catalytic parameters for V-MoS$_2$ in terms of the V concentration and the substrate: onset potential, overpotential at 10 mA cm$^{-2}$ ($\eta_{10}$), exchange current density ($j_0$), Tafel slope, double layer capacitance ($C_{dl}$), relative electrochemically active surface area (EASA) extracted with respect to the substrate and charge transfer resistance ($R_{ct}$).

| Sample | Onset Potential (mV) | $\eta_{10}$ (mV) | $j_0$ (mA cm$^{-2}$) | Tafel Slope (mV dec$^{-1}$) | $C_{dl}$ (mF cm$^{-2}$) | Relative EASA | $R_{ct}$ ($\Omega$) |
|---|---|---|---|---|---|---|---|
| V$_{(0.8\%)}$-MoS$_2$/Gr | -313 | -444 | $2.00 \times 10^{-2}$ | 176 | 15.3 | 1.18 | 5 |
| V$_{(1.6\%)}$-MoS$_2$/Gr | -244 | -413 | $2.00 \times 10^{-2}$ | 135 | 18.4 | 1.42 | 4.5 |
| V$_{(7.3\%)}$-MoS$_2$/Gr | -103 | -206 | 0.143 | 82.1 | 24.6 | 1.89 | 3.0 |
| V$_{(9.3\%)}$-MoS$_2$/Gr | -52 | -120 | 0.283 | 54.2 | 28.2 | 2.17 | 3.0 |
| V$_{(9.3\%)}$-MoS$_2$/Ni | -30 | -99 | 0.380 | 46 | 30.1 | 2.46 | 2.0 |
| V$_{(9.3\%)}$-MoS$_2$/Cu | -10 | -60 | 0.650 | 40 | 35.3 | 2.73 | 0.8 |
| Pristine MoS$_2$/Cu | -208 | -280 | $2.00 \times 10^{-3}$ | 130 | 16.2 | 1.17 | 6.0 |
| Pristine MoS$_2$/Ni | -225 | -293 | $1.70 \times 10^{-3}$ | 132 | 14.8 | 1.21 | 7.6 |
| Pristine MoS$_2$/Gr | -323 | -483 | $1.20 \times 10^{-3}$ | 211 | 13.2 | 1.02 | 11.1 |
| Bare Cu | -80 | -189 | 0.139 | 59.1 | 13.8 | 1.00 | 5.0 |
| Bare Ni | -130 | -270 | 0.102 | 84.0 | 12.2 | 1.00 | 6.0 |
| Bare Gr | -353 | -504 | $1.00 \times 10^{-3}$ | 213 | 13.0 | 1.00 | 9.8 |
| Pt/C | -5 | -21 | 0.800 | 32 | - | - | 2.2 |

**Table S3.** Comparison of onset potential (onset potential$_{normalized}$) and overpotential at 10 mA/cm$^2$ ($\eta_{10normalized}$) after EASA normalization.

| Material | onset potential$_{normalized}$ (mV) | $\eta_{10normalized}$ (mV) | Reference |
|---|---|---|---|
| V$_{(9.3\%)}$-MoS$_2$ | -20 | -100 | Our work |
| MoS$_2$ | -53 | -171 | 18 |
| Co$_2$P@CP | -60 | -140 | 19 |
| Ni-W$_2$C | -50 | -300 | 20 |
| Ni-Pt film | -80 | -250 | 21 |

**Table S4.** Exchange current density ($j_0$) contribution by $vac_s$ in pristine and $V_{(9.3\%)}$-MoS$_2$.

| Material | $J_0$ (mA cm$^{-2}$) | % Mo-vac$_s$ | % V-vac$_s$ |
|---|---|---|---|
| Pristine MoS$_2$ | $2.0 \times 10^{-3}$ | 2.1 | 0 |
| $V_{(9.3\%)}$-MoS$_2$ | 0.65 | 0.8 | 1.7 |